\documentclass[12pt]{article}
\usepackage{graphicx}
\usepackage{tabularx, booktabs}
\newcolumntype{Y}{>{\centering\arraybackslash}X}
\newcolumntype{B}{>{\centering\arraybackslash\hsize=1.5\hsize}X}
\newcolumntype{Q}{>{\centering\arraybackslash\hsize=2\hsize}X}
\usepackage{titlesec}
\usepackage{wrapfig}
\usepackage{epstopdf}
\usepackage{setspace}
\usepackage[left=1in,top=1in,right=1in,bottom=1in,nohead,paperwidth=8.5in, paperheight=11in]{geometry}

\usepackage{amsmath}
\usepackage[title,titletoc,toc]{appendix}
\usepackage{color}
\usepackage{tikz}
\usepackage{subcaption}
\usepackage{caption}

\usepackage[round]{natbib}
\usepackage{lipsum}
\usepackage{indentfirst}
\usepackage{hyperref}          
\hypersetup{
	colorlinks=true,
	citecolor=blue
}
\graphicspath{ {Figures/} }

\usepackage{dsfont}

\usepackage{algorithm}
\usepackage{algorithmic}
\usepackage{leading} 

\usepackage{setspace}

\title{\vspace{-1.0cm}  
	\bf MLOps Monitoring at Scale for Digital Platforms\footnote{Correspondence to 
		Ines Wilms, Maastricht University, i.wilms@maastrichtuniversity.nl. 
		Yu Jeffrey Hu, Purdue University, yuhu@purdue.edu. 
		Jeroen Rombouts, Essec Business School. We are very grateful to Benjamin Wolter, Pablo Perez Piskunow and Roger Caminal for expert advice, and Olivier Scaillet for comments and checks provided on earlier versions of the paper. IW was financially supported by the Dutch Research Council (NWO) under grant number VI.Vidi.211.032.}}

\author{
	Yu Jeffrey Hu,  Jeroen Rombouts and  Ines Wilms 
}

\begin{document}
	
	\begin{titlepage}
		\clearpage\thispagestyle{empty}
		\maketitle
		
		\begin{singlespace}
			\noindent
			{\bf Abstract.} 
			Machine learning models are widely recognized for their strong performance in forecasting.
			To keep that performance in streaming data settings, they have to be monitored and frequently re-trained. This can be done with machine learning operations (MLOps) techniques under supervision of an MLOps engineer. However, in digital platform settings where the number of data streams is typically large and unstable, 
			standard monitoring becomes  either suboptimal or too labor intensive for the MLOps engineer. As a consequence,  companies often fall back on very simple worse performing ML models without monitoring. We solve this problem by adopting a design science approach and introducing a new monitoring framework, the  Machine Learning Monitoring Agent (MLMA), that is designed to work at scale for any ML model with reasonable labor cost. A key feature of our framework concerns test-based automated re-training based on a data-adaptive reference loss batch.
			The MLOps engineer is
			kept in the loop via key metrics and also acts, pro-actively or retrospectively, to maintain performance of the ML model in the production stage. We  conduct a large-scale  test at a last-mile delivery platform to empirically validate our monitoring framework. 
		\end{singlespace}

		
		\thispagestyle{empty}
	\end{titlepage}

	\doublespacing

	\clearpage
	
	\section{Introduction} \label{Introduction}
	
	The recent advances in artificial intelligence (AI) may be as transformative to the economy as the industrial revolution \citep{abis2024changing}. While some applications of the technology catch worldwide news coverage,  the critical question and challenge for companies today is ``How to distribute work between humans and AI to increase productivity?".  
	In this paper, we focus on digital platforms which are well-suited for benefiting from AI for productivity gains at large scale. In fact, the products (e.g. deliveries in our application) offered by a platform are available thanks to a carefully designed process which relies on Machine Learning Operations (MLOps) involving highly skilled workers who are scarce on the job-market. The goal of this paper is to facilitate the creation of productivity gains for MLOps engineers by offloading cornerstone MLOps monitoring tasks to AI. This allows platforms to solve their problem of managing advanced algorithms at scale.

	MLOps is an iterative process that integrates machine learning development and operations to automate, deploy, and monitor ML models in production
	\citep{testi2022mlops, kreuzberger2023machine}. MLOps emerged from the convergence of the machine learning, software development operations, and data engineering disciplines. While academic research in the machine learning (ML) community concentrates on building AI/ML models, and the information systems (IS) community has seen a proliferation of research on managing AI and the benefits and pitfalls of human collaboration with machines, academic work on MLOps is still in its infancy. This paper addresses an essential part of MLOps, namely continuous monitoring. The main objective of continuous monitoring is risk management of in-production models by checking for performance drift;  the foundation of any robust ML-based product \citep{testi2022mlops, ruf2021demystifying}.
	
	In particular, we focus on MLOps monitoring at digital platforms since there is both strong potential and need for further automation of this task. The potential is large since digital platforms inherently operate on an AI core (e.g., \citealp{rai2019next}),  they constitute a significant part of the economy and host a substantial share of the working population \citep{zgola21, wilson2023}. Improving the functioning of digital platforms, and more generally any business striving for process and workflow automation, would therefore translate to a sizable impact. 
	Also, the need for further automation of the monitoring task is large. Platforms are, from an analytics viewpoint, complex prescriptive systems which heavily rely on forecasts. Indeed, a key success factor for operating two-sided platforms concerns their ability to accurately forecast demand and supply-- indispensable inputs when optimizing work planning and compensation schemes --in a rapidly changing environment with volatile market conditions due to competition, events and changing customer preferences \citep{hu2023fast}. Platform data are ``streaming” as they are generated continuously by events, typically with new batches entering at high velocity-- 15-min intervals in our application --and high granularity. In fact, platforms typically operate in thousands of verticals such as delivery areas, product categories. Another unique feature of platform data streams is that they are subject to frequent shifts. Such instabilities cause forecast breakdowns.
	This requires fast forecasting, monitoring of forecast performance at scale, and model re-training when new data streams arrive.

	ML models have proven to be attractive tools to deliver forecasts in such challenging environments (see various forecast competitions, e.g., \citealp{bojer2021kaggle, makridakis2022m5}). 
	Once the data scientist has done the relevant feature engineering, sophisticated off-the-shelf implementations are routinely available thereby facilitating their continuous deployment in production environments. 
	The MLOps engineer then monitors the ML infrastructure and manages the automated ML workflow pipelines \citep{kreuzberger2023machine}. When doing so, (s)he
	faces the problem that an ML model in production runs a non-negligible risk that new input data is fundamentally different from the training data, therefore potentially producing large forecast errors if the model is not sequentially re-trained.  This calls for tools to monitor the forecast performance of in-production models that, in turn, allow the MLOps engineer to provide feedback to the scheduler who enables re-training of the ML model; an MLOps topic on which there is a dearth of research.  In this paper, we therefore address the following concrete question: ``How to monitor forecast performance such that re-training of an in-production ML model is  triggered to avoid systematic deterioration of forecast performance?"
	
	There are at least four strategies for monitoring the need to re-train a large scale forecasting system: i) do nothing, ii) periodic re-training, iii) on demand and iv) our proposal. The first strategy is no monitoring since re-training of the model is not needed. An example of such a system is taking the last available observation as a forecast. Such procedures, however, typically deliver forecasts that are easily dominated by slightly more sophisticated algorithms such as random forests. Even though it seems obvious that this is not a good strategy, it is frequently used by companies, hence we consider it as a monitoring benchmark.
	The second strategy is periodic and automatic re-training either upon arrival of every new data batch or at pre-specified dates (every six months for example). The former is, however, resource-heavy (namely labor and computing time) and unfeasible in high-frequency settings (like the one in our application). The latter is computationally less costly (if re-training is infrequent) and better than doing nothing, but is likely to be sub-optimal because forecast breakdowns do not necessarily occur simultaneously across all components of the system and they can occur at any date between the scheduled re-training dates.
	The third strategy is monitoring on demand, for example by setting a performance threshold that triggers the re-training of the ML model. This can also require expert input from an MLOps engineer or from the business which takes up costly time to decide when to re-train. Besides, work by \cite{kremer2011demand} indicates that humans typically overreact to forecast losses which would entail unnecessary re-training.  
	The fourth strategy is the one we propose.
	
	This paper takes a design science approach as defined by \cite{Hevner_March_Park_Ram_2004}; we propose the Machine Learning Monitoring Agent (MLMA),  a monitoring strategy for when to re-train in-production ML models. It rests on the key premise that MLOps engineers are scare and expensive resources that are difficult to recruit and retain. Our monitoring framework is data-driven, can operate in a computationally-efficient manner with minimal human intervention, yet still keeps the MLOps engineer in the loop via well-defined interpretable metrics. The re-training decision is thus made by combining algorithmic automation with information available only to the human.  In addition to productivity gains (in terms of labor and computing costs), the framework also allows  forecast accuracy gains, thanks to the use of monitored ML models, compared to standard forecast approaches.

	As a first methodological contribution, we offer an alert system based on a standard statistical test that is inspired on the work of \cite{Luo_JASA_2022_batches}. 
	The framework triggers statistical test-based alerts for re-training of the ML model under consideration when forecast performance-- as measured by a streaming loss function-- changes compared to a well-defined reference batch that is updated at each date a forecast performance shift is detected. Re-training is thus done at a priori unknown times and only when deemed necessary to avoid forecast deterioration due to a changing environment, thereby ensuring a good balance between labor and computing costs, and forecast accuracy. A key component of MLMA is that it operates with a data-driven streaming performance reference batch, which is different from an on-demand re-training strategy that works with fixed performance thresholds; the levels of which are set on an ad-hoc basis.
	By enhancing the automation of the ML model re-training cycle, the MLOps engineer receives only alerts in case of ML model deterioration and can then subsequently decide to re-train or not. This frees up time for them to monitor the frequency as well as other metrics related to the re-training of in-production ML models. Finally, note that by using a standard statistical testing procedure, we monitor an interpretable metric and thereby reduce complexity to a maximum. This facilitates adoption by the MLOps engineer and portability across IT systems, see e.g.,  \cite{das2024drives}, thereby avoiding the need to rely on advanced data science packages and permitting the MLOps engineer to explain results and provide feedback in simple terms to business colleagues.

	As a second main contribution, we assess the utility of MLMA in a business environment, which aligns with the core principles of design science.
	Indeed, to fully mimic the complex environment of the digital platform (e.g., \citealp{Benbya2020}) with new incoming data batches, re-training, data and performance drifts, it is essential to study improvements to MLOps monitoring in an actual real-world production setting as opposed to a sandbox setting used frequently in academic research. On the other hand, practitioners are currently using several best practices retrieved from software vendors documentation but the actual value of which is unknown given the lack of a formal comparison between existing monitoring strategies. Our paper is the first to address this issue by quantifying the value that different monitoring strategies bring to the table. We demonstrate the advantages of our MLMA MLOps monitoring framework by offering an example of an on-demand delivery platform operating in London. The platform faces 15-min demand (January 2019 to March 2021) across  the 32  areas that constitute its London market. It produces demand forecasts using a random forest model that is constructed with domain expertise; building models with such expertise is a successful practice as shown in \cite{ibrahim2021eliciting}. In our paper, we take this established ML model to forecast demand for its 32 delivery areas as input and assess our proposed monitoring strategy vis-à-vis the other discussed, popular monitoring strategies.

	Performance of monitoring strategies is often assessed solely through statistical measures of forecast error, overlooking the potentially massive computational costs—including processing time and environmental impact—that come with generating the forecasts.
	In this paper, we compare performance in terms of forecast accuracy, labor salary and computing/CO2 costs. In terms of forecast accuracy, our MLMA framework yields more accurate forecasts compared to its benchmarks. The only strategy that obtains higher accuracy, namely daily re-training, is too costly in labor and computing for any platform business.
	In terms of productivity gains, we find that our data-driven framework is four times faster than its manual variant. The implied computing costs are fairly low and controllable by the MLOps engineer.
	In terms of insights, apart from a better business understanding on the stability of commercial activity in local areas, our framework generates interpretable metrics such that MLOps engineers can monitor the durations between re-training days, and the predictive power of the features used in the ML model. These metrics can serve as valuable input to the MLOps engineer when providing feedback from the monitoring component to the model training infrastructure of the MLOps system.
	Finally, though we demonstrate the applicability of the MLMA for on-demand platform forecasting, it can be used in a wide range of other applications that necessitate monitoring of streaming forecasts subject to change; we provide an overview of such opportunities at the end of the paper.
	
	The rest of the paper is organized as follows.
	Section \ref{Literature} positions our contribution in the relevant IS, computer science and statistics literature. 
	Section \ref{Methodology} presents the new MLMA framework to monitor forecast performance which triggers  re-training of a given in-production ML model. 
	Section \ref{PlatformBusiness} introduces the platform business problem and the data we analyze. It also describes the forecast setup and benchmark procedures used in the application. 
	Section \ref{sec:forecast_results} presents the monitoring framework results by reporting on forecast accuracy, labor and computing costs.
	Section \ref{sec:implications} provides business implications and MLOps engineer insights generated by MLMA.
	Finally, Section \ref{Conclusion} concludes the article, suggests future research directions and discusses general applicability for other data streaming applications.

	\section{Literature Review} \label{Literature}
	
	Our paper is related to various disciplines; we briefly review related literature on managing AI (Section \ref{lit:mngmt}), MLOps (Section \ref{lit:mlops}) and monitoring streaming data (Section \ref{lit:monitor}).

	\subsection{Managing Human-Machine Collaborations} \label{lit:mngmt}
	Our paper contributes to the emerging management and IS literature on human-machine collaborations.
	\cite{fugener2021will} discuss merits and pitfalls for humans working with AI, \cite{Sturm2021} study how organizations can coordinate human and machine learning to ensure their harmonious functioning and
	\cite{Boyaci_mansci_2024}  investigate human-machine collaboration in a rational inattention framework to stipulate environments where their collaboration is likely to be most beneficial.
	\cite{Stelmaszak2025} introduce the concept of distributed delegation and examine how algorithms delegate tasks to multiple humans in Uber’s ride-hailing platform, by analyzing patent data and interviews with drivers and passengers. They highlight the hybrid and collective nature of algorithmic delegation and challenge full algorithmic autonomy. \cite{Balakrishnan_2025} find that humans tend to naively average their own predictions with algorithmic recommendations, often misusing private information. They show that feature transparency and targeted interventions help users make better adjustments, and therefore improve prediction accuracy.
	
	Other studies investigate the value of keeping the ``human-in-the-loop" as compared to full machine automation. 
	Amogst others, 
	\cite{Fugener_ISR_2022}
	demonstrate through experimental studies on image classification how human-AI collaboration is beneficial only when the AI delegates work to humans, while
	\cite{Tian_ISR_2024}
	highlight value of human involvement  in microloan field experiments  only when extensive information was coupled with machine explanations.
	
	Finally, our paper also contributes to the line of research on machine-based prescriptions and recommendations. 
	\cite{Sun_mansci_2022} 
	study non-conforming human behavior with algorithmic prescriptions through a large-scale field experiment at warehouses of the Alibaba Group and highlight the importance of integrating
	behavioral components
	to improve performance.
	\cite{de2023your} investigate the extent to which decision makers can properly assess machine-based recommendations, thereby offering guidelines on the decision to adopt or reject machine prescriptions.
	\cite{Wang_mansci_2024} 
	study the effect of experience on the collaboration dynamics between humans and AI for medical chart coding and find evidence that AI  mainly benefits junior workers with greater task-based experience more so than senior workers with lower trust in AI.
	
	\subsection{MLOps} \label{lit:mlops}
	Although MLOps is a well-established term in the industry, the academic computer science literature on MLOps is still in its infancy. The management and IS literature on MLOps is to the best of our knowledge nonexistent, a gap that this paper fills. We refer the reader to the excellent recent overviews by \cite{testi2022mlops} and \cite{kreuzberger2023machine}; and references therein.
	The former focuses on defining clear, standardized methodologies and operations for conducting MLOps projects,
	whereas the latter offers an aggregated overview of four main MLOps aspects, namely its principles, components, roles and architecture.
	
	An important MLOps topic in the context of our work that did get increasingly more attention in the ML literature is related to concept drift, which occurs in real-world applications when  input data changes over time, see e.g., \cite{suarez2022survey} for a survey.  Detecting concept drifts as feedback mechanism is crucial to enable periodic re-training of the ML model based on new input data;  as part of the  MLOps principle on ``Continuous ML training and evaluation" \citep{kreuzberger2023machine}.
	Methodology-wise, 
	theoretical and practical frameworks emerge in the literature that study how to monitor and deal with such concept drifts, see e.g.,  \cite{AHMAD2017134_Neurocomputing, Rabanser_neurips_2019, Federici_neurips_2021, Zhang_neurips_2021}. 
	
	This paper differs from the above because our monitoring scheme is aimed at detecting forecast deterioration, i.e.\ \textit{a shift in forecast performance} (aka ``performance drift"; \citealp{testi2022mlops}) rather than data shifts; hence it is geared towards the MLOps principle on ``continuous monitoring" \citep{kreuzberger2023machine} which forms the foundation of a robust ML-based product \citep{ruf2021demystifying}.
	Indeed, when forecasting demand at delivery platforms, the instability in the underlying dynamics of the data streams is of second order importance;  what matters is the  forecast performance stability of the ML model.
	The focus on forecast performance simplifies the monitoring procedure (one forecast loss stream versus multiple data streams) and makes it easy to understand (bad forecasts can have severe business implications); thereby offering strong potential for automation.
	Yet  the operationalization of this continuous monitoring task has been largely overlooked as compared to the task of monitoring data shifts; a gap that this paper fills.
	
	Lastly, given that many companies are training staff of entire business units to become ``data citizens", it is crucial for adoption that such a continuous monitoring procedure is easy to explain, implement,  fast to run and portable \citep{das2024drives}. Our proposal offers these appealing features.
	Before detailing the proposed monitoring procedure in Section \ref{Methodology}, the next section presents related work on monitoring data streams and forecast performance.
	
	\subsection{Forecast Instability and Monitoring Data Streams} \label{lit:monitor}
	Our work is related to the econometrics literature on forecasting in presence of instabilities as well as the statistics literature on monitoring data streams.
	In econometrics, \cite{Giacomini_Rossi_2009} have pioneered the literature on ``forecast breakdown", thereby proposing forecast deterioration tests; see \cite{Rossi_JEL_2021} for a  survey.
	These methods are, however, typically retrospective and hence not suitable in continuously evolving streaming data settings, which is the object of interest in this paper.
	
	In statistics, there is a vast literature on monitoring compatibility of an estimated model when new data becomes available. This research dates back to \cite{White_ECA_1996}  who consider real-time monitoring for linear regression models, with recent extensions to 
	multivariate data streams in e.g., \cite{Mei_2010, 
		Dette_JASA_2020}.
	Another closely related line of statistical research consists of monitoring data streams by sequentially testing a null hypothesis of absence of anomalies, see e.g., \cite{Ross_2011_Technometrics, Gang_2023}.
	Of particular importance here is the work by 
	\cite{Luo_JASA_2022_batches}, who monitor abnormal data batches in data streams, as our monitoring proposal is inspired on their work.
	Yet, these methods are geared towards monitoring streaming data, not towards streaming forecast loss streams which are needed for our monitoring proposal. Since forecast loss streams are highly nonlinear functions of the underlying data streams, dependent, and their inherent nonstationarity yields frequent forecast performance shifts, we extend the approach of \cite{Luo_JASA_2022_batches}, originally aimed at monitoring abnormal data batches.
	
	\section{MLOps Monitoring Framework} \label{Methodology}
	In Section \ref{subsec:outline_flow}, we give the outline of the new monitoring framework we propose  using a flow diagram.
	Section \ref{subsec:problem}  introduces the problem of monitoring forecasting performance with streaming data.
	In Section  \ref{subsec:monitoring}, we  introduce our automated monitoring procedure, and explain 
	in Section \ref{subsec:monitoring_productivity} how the MLOps engineer is kept in the loop.
	Finally, Section \ref{subsec:algorithm} details the algorithmic view of our framework.
	
	\subsection{Outline of Framework} \label{subsec:outline_flow}
	We propose a new test-based automated framework to monitor forecasting performance of in-production ML models while keeping the MLOps engineer in the loop. We name our framework MLMA, Machine Learning Monitoring Agent, to reflect this overall objective.
	
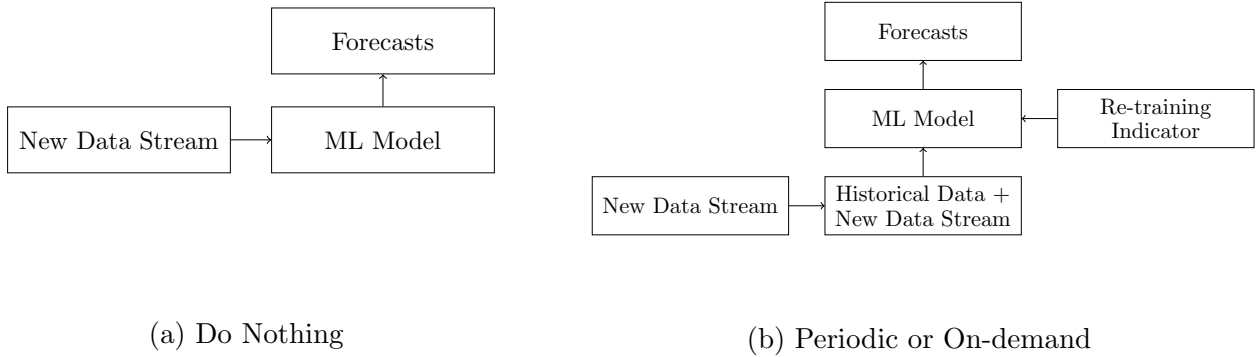
\begin{figure}[t]
		\begin{minipage}{0.4\textwidth}
			\resizebox{\textwidth}{!}{ 
				\begin{tikzpicture} 
					\node[draw, rectangle, text width=3.1cm, minimum width = 3cm, minimum height = 1cm, align=center, font=\fontsize{11}{12}\selectfont] (A) at (1,0) [draw, rectangle] {New Data Stream};
					\node[draw, rectangle, text width=3.1cm, minimum width = 3cm, minimum height = 1cm, align=center, font=\fontsize{11}{12}\selectfont] (B) at (5,0) [draw, rectangle] {ML Model};
					\node[draw, rectangle, text width=3.1cm, minimum width = 3cm, minimum height = 1cm, align=center, font=\fontsize{11}{12}\selectfont] (C) at (5, 1.5) [draw, rectangle] {Forecasts};
					
					\path (A.east) edge[->] (B.west);
					\path (B.north) edge[->] (C.south);
			\end{tikzpicture}}
			\vspace{1.2cm}
			\subcaption{Do Nothing}      
		\end{minipage} 
		\hspace{1cm}
		\begin{minipage}{0.54\textwidth}   
			\resizebox{\textwidth}{!}{    
				\begin{tikzpicture}
					\node[draw, rectangle, text width=3.1cm, minimum width = 3cm, minimum height = 1cm, align=center, font=\fontsize{11}{12}\selectfont] (A2) at (1,0) [draw, rectangle] {New Data Stream};
					\node[draw, rectangle, text width=3.1cm, minimum width = 3cm, minimum height = 1cm, align=center, font=\fontsize{11}{12}\selectfont] (B2) at (5,0) {Historical Data + New Data Stream};
					\node[draw, rectangle, text width=3.1cm, minimum width = 3cm, minimum height = 1cm, align=center, font=\fontsize{11}{12}\selectfont] (C2) at (5,1.5) [draw, rectangle] {ML Model};
					\node[draw, rectangle, text width=3.1cm, minimum width = 3cm, minimum height = 1cm, align=center, font=\fontsize{11}{12}\selectfont] (D2) at (5,3) [draw, rectangle] {Forecasts};
					\node[draw, rectangle, text width=3.1cm, minimum width = 3cm, minimum height = 1cm, align=center, font=\fontsize{11}{12}\selectfont] (E2) at (9,1.5) [draw, rectangle] {Re-training Indicator};
					
					\path (A2.east) edge[->] (B2.west);
					\path (B2.north) edge[->] (C2.south);
					\path (C2.north) edge[->] (D2.south);
					\path (E2.west) edge[->] (C2.east);
			\end{tikzpicture}}
			\vspace{0.45cm}
			\subcaption{Periodic or On-demand}
		\end{minipage}
		\caption{Existing Monitoring Strategies.} \label{fig:monitoring_existing}
	\end{figure}

	Before we detail our framework, it is useful to visualize the existing monitoring strategies discussed in the introduction. Figure \ref{fig:monitoring_existing} displays on the left the ``Do nothing" strategy. Given an initially trained model or initial data batch, each time a new data stream comes in, it is used for producing new forecasts but not for re-training the ML model. The ``Periodic' or ``On demand" monitoring strategies displayed on the right  of Figure \ref{fig:monitoring_existing} differ in two aspects. First, new data streams and historical data are gathered together, hence all information is stored. Second, a re-training indicator signals when re-training with all available data is required. In the periodic monitoring strategy, this happens at fixed time intervals such as each semester. In the on-demand monitoring strategy, this occurs for example when the business anticipates upcoming instability such as specific local events, or when the accuracy of the models is too low to satisfy business needs.
	
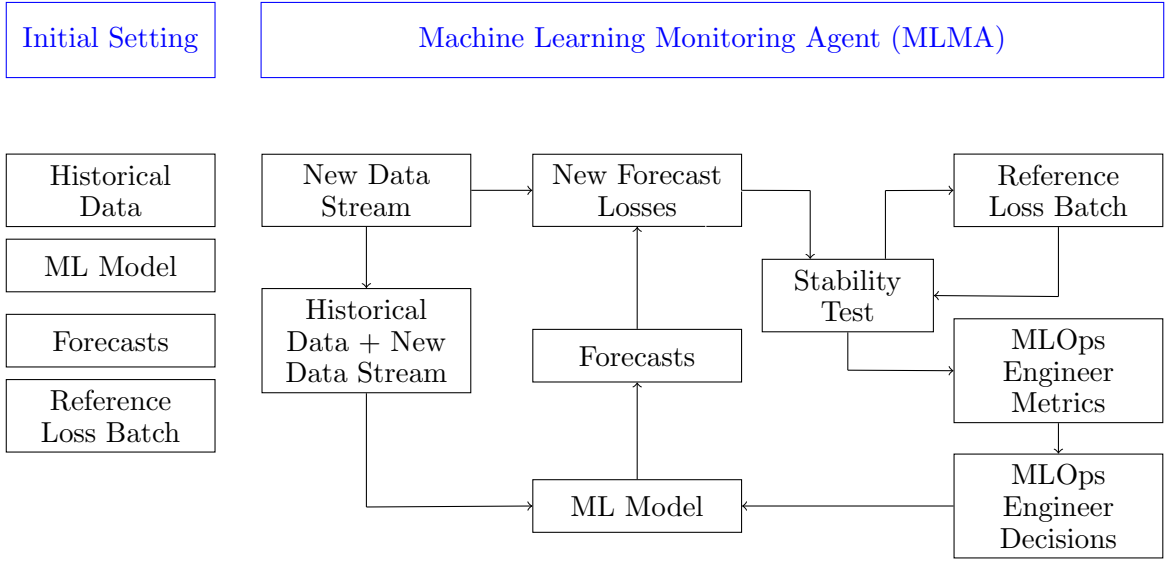
\begin{figure}[t]
		\centering
		\begin{tikzpicture}
			\node[draw=blue, text=blue, rectangle, text width=2.5cm, minimum width = 2.5cm, minimum height = 1cm, align=center, font=\fontsize{11}{12}\selectfont] (A) at (0,12) [draw, rectangle] {Initial Setting};
			
			\node[draw, rectangle, text width=2.5cm, minimum width = 2.5cm, minimum height = 0.7cm, align=center, font=\fontsize{11}{12}\selectfont] (B) at (0,10) [draw, rectangle] {Historical Data};
			\node[draw, rectangle, text width=2.5cm, minimum width = 2.5cm, minimum height = 0.7cm, align=center, font=\fontsize{11}{12}\selectfont] (C) at (0,9) [draw, rectangle] {ML Model};
			\node[draw, rectangle, text width=2.5cm, minimum width = 2.5cm, minimum height = 0.7cm, align=center, font=\fontsize{11}{12}\selectfont] (D) at (0,8) [draw, rectangle] {Forecasts};
			\node[draw, rectangle, text width=2.5cm, minimum width = 2.5cm, minimum height = 0.7cm, align=center, font=\fontsize{11}{12}\selectfont] (E) at (0,7) [draw, rectangle] {Reference Loss Batch};
			
			\node[draw=blue, text=blue, rectangle, text width=10cm, minimum width = 12cm, minimum height = 1cm, align=center, font=\fontsize{11}{12}\selectfont] (F) at (8,12) [draw, rectangle] {Machine Learning Monitoring Agent (MLMA)};
			
			\node[draw, rectangle, text width=2.5cm, minimum width = 2.5cm, minimum height = 0.7cm, align=center, font=\fontsize{11}{12}\selectfont] (G) at (3.4,10) [draw, rectangle] {New Data Stream};
			
			\node[draw, rectangle, text width=2.5cm, minimum width = 2.5cm, minimum height = 0.7cm, align=center, font=\fontsize{11}{12}\selectfont] (H) at (3.4,8) [draw, rectangle] {Historical Data + New Data Stream};
			
			\node[draw, rectangle, text width=2.5cm, minimum width = 2.5cm, minimum height = 0.7cm, align=center, font=\fontsize{11}{12}\selectfont] (I) at (7, 5.8) [draw, rectangle] {ML Model};
			\node[draw, rectangle, text width=2.5cm, minimum width = 2.5cm, minimum height = 0.7cm, align=center, font=\fontsize{11}{12}\selectfont] (J) at (7,7.8) [draw, rectangle] {Forecasts};
			\node[draw, rectangle, text width=2.5cm, minimum width = 2.5cm, minimum height = 0.7cm, align=center, font=\fontsize{11}{12}\selectfont] (K) at (7,10) [draw, rectangle] {New Forecast Losses};
			
			\node[draw=white, rectangle, text width=2.5cm, minimum width = 2.5cm, minimum height = 0.7cm, align=center, font=\fontsize{11}{12}\selectfont] (invisibleHI) at (3.4, 5.43) [draw, rectangle] { };
			\node[draw=white, rectangle, text width=2.5cm, minimum width = 2.5cm, minimum height = 0.7cm, align=center, font=\fontsize{11}{12}\selectfont] (invisibleKL) at (9.3, 9.64) [draw, rectangle] { };
			\node[draw=white, rectangle, text width=2.5cm, minimum width = 2.5cm, minimum height = 0.7cm, align=center, font=\fontsize{11}{12}\selectfont] (invisibleLM) at (10.3, 9.64) [draw, rectangle] { };
			
			\node[draw=white, rectangle, text width=2.5cm, minimum width = 2.5cm, minimum height = 0.7cm, align=center, font=\fontsize{11}{12}\selectfont] (invisibleLN) at (9.8, 7.25) [draw, rectangle] { };
			
			\node[draw=white, rectangle, text width=2.5cm, minimum width = 2.5cm, minimum height = 0.7cm, align=center, font=\fontsize{11}{12}\selectfont] (invisibleLM2) at (12.6, 8.23) [draw, rectangle] { };
			
			\node[draw, rectangle, text width=2cm, minimum width = 2cm, minimum height = 0.7cm, align=center, font=\fontsize{11}{12}\selectfont] (L) at (9.8,8.6) [draw, rectangle] {Stability Test};
			\coordinate (L_shiftedL) at ([xshift=-0.5cm] L.north);
			\coordinate (L_shiftedR) at ([xshift=0.5cm] L.north);
			
			\node[draw, rectangle, text width=2.5cm, minimum width = 2.5cm, minimum height = 0.7cm, align=center, font=\fontsize{11}{12}\selectfont] (M) at (12.6,10) [draw, rectangle] {Reference Loss Batch};
			
			\node[draw, rectangle, text width=2.5cm, minimum width = 2.5cm, minimum height = 0.7cm, align=center, font=\fontsize{11}{12}\selectfont] (N) at (12.6, 7.6) [draw, rectangle] {MLOps Engineer Metrics};
			
			\node[draw, rectangle, text width=2.5cm, minimum width = 2.5cm, minimum height = 0.7cm, align=center, font=\fontsize{11}{12}\selectfont] (O) at (12.6, 5.8) [draw, rectangle] {MLOps Engineer Decisions};
			
			\path (G.south) edge[->] (H.north);
			\path (H.south) edge[-] (invisibleHI);
			\path (invisibleHI.north) edge[->] (I.west);
			\path (I.north) edge[->] (J.south);
			\path (J.north) edge[->] (K.south);
			\path (G.east) edge[->] (K.west);
			
			\path (K.east) edge[-] (invisibleKL.north);
			\path (invisibleKL.north) edge[->] (L_shiftedL);
			\path (L_shiftedR) edge[-] (invisibleLM.north);
			\path (invisibleLM.north) edge[->] (M.west);
			\path (invisibleLM2.north) edge[->] (L.east);
			\path (invisibleLN.north) edge[-] (L.south);
			\path (invisibleLN.north) edge[->] (N.west);
			
			\path (M.south) edge[-] (invisibleLM2.north);
			
			\path (N.south) edge[->] (O.north);
			\path (O.west) edge[->] (I.east);
			
		\end{tikzpicture}  
		\caption{MLMA Framework Outline.} \label{fig:MLOps_framework}
	\end{figure}

	Figure \ref{fig:MLOps_framework} visualises the MLMA framework in a diagram. Since we monitor in-production ML models, the initial setting of the production environment is given. It provides the following important pieces of information for the later monitoring stage:
	A historical data batch, a selected ML model that is trained and that produces a first set of forecasts for which forecast losses, based on forecast errors being  the difference between actuals and forecasts,  can be computed
	and collected in a reference loss batch. 
	Next, the MLMA is activated upon arrival of every new data stream, which serves two purposes.
	First, this batch contains ground truths (``actuals") for earlier made forecasts, and hence are used to compute new forecast losses. Second, this batch is added to the historical data stream, which is needed to feed the actual ML model to produce new forecasts. 
	The stability test then determines whether re-training of the ML model is recommended or not. To this end, it uses the new forecast losses which are compared against the reference loss batch.
	The outcome of this statistical test impacts the process in two ways. First, if there is forecast stability (when the new forecast losses are alike the reference loss batch) then the reference loss batch is augmented with the new forecast losses and no re-training of  the ML model is needed. In case of instability (when the new forecast losses significantly differ from the reference loss batch), the new forecast losses become the new reference batch. Second, a re-training alert is communicated to the MLOps engineer who decides if the ML model is re-trained in case of this detected instability case. The MLOps engineer can also always take action based on external information. 
	
	The MLMA framework differs in a fundamental way from an automated on-demand re-training strategy for two main reasons. First, MLMA works with a reference batch that is data-driven and dynamically adapting according to the statistical properties of forecast loss streams. In contrast, on-demand re-training fixes a priori a threshold based on which re-training is then decided. Such a strategy requires pre-setting the threshold, which is context specific and which is not allowed to be adaptable in a setting of unstable data streams. Second, MLMA fosters human-algorithm collaboration by actively keeping the MLOps engineer in the loop. This allows humans to pro-actively decide on the 
	re-training of the ML model based on additional information outside the historical data streams that may arrive on the fly. Furthermore, having an MLOps engineer in the loop enables a thorough, expert-based assessment of whether the ML model is at risk and requires a more substantial revision beyond simply re-training with new data. MLMA supports this important feedback loop in the MLOps architecture to better inform data scientists in charge of making ML model refinements.
	In an on-demand re-training strategy, the MLOps passively re-trains ML models and risks missing out on relevant information regarding the model's performance.
	
	\subsection{Technical Setup} \label{subsec:problem}
	Let  $d_t = \left( d_{1t}, d_{2t}, \ldots, d_{Dt}  \right)$ 
	denote the $D$-variate data stream for $t \ge 1$.  The data is non-identically distributed and dependent across time and streams. 
	Data streams arrive in batches of fixed size $B$, i.e.\
	the number of observations that simultaneously enter the data streams. 
	An ML model is needed to produce direct forecasts for horizons $q=1, \ldots, Q$. 
	In our platform application, 
	we have demand data  $d_{it}$ on $i=1,\ldots D=32$ boroughs for each quarter hour $t$, the batch size is $B=60$ since each business day 60 new demand values arrive for each of the quarter hours of the 15 business hours.
	Forecasts are needed at the end of each business day, and this for each of the $Q=60$ quarter hours of the next business day.\footnote{For ease of notation, we take the batch size and forecast horizon equal, but these can differ in applications.}
	
	We model each data stream $d_{it}$ separately conditional on all historical data streams $d_{t-j}$ for $j>1$ (as opposed to $d_{t}$ jointly conditional on $d_{t-j}$) since the monitoring procedure and corresponding re-training scheme is likely to be stream-specific.
	A general ML model for data stream $i$ at time $t$ that uses all past data streams as well as other relevant information such as deterministic trends and seasonal effects, denoted by $a_t$, can be written  as
	\begin{eqnarray}
		d_{it} = f_{it}\left(\{d_{t-j}\}_{j \in \mathcal{J}}, a_t\right)  + \varepsilon_{it}, \label{eq:generalmodel}
	\end{eqnarray}
	where $f_{it}(\cdot)$ denotes the ML model  
	for stream $i$ at time $t$ and $\varepsilon_{it}$ is an additive error term.
	
	We explicitly index the ML model $f_{it}(\cdot)$ with $t$ to stress that our setting is nonstandard compared to the classical framework of estimating time invariant functions $f_{i}(\cdot)$. In the latter case, theory is available for both parametric and nonparametric estimators $\hat f_{i}$ that guarantee   convergence  to $f_{i}$ for increasing sample size.\footnote{ \cite{Delaigle_2023} propose kernel estimators for the density of data streams which smoothly vary over time. 
		For streaming data and constant parametric functions, \cite{Luo_2020_renewable} show that renewable estimators have similar properties when the number of data batches goes to infinity.} However, 
	these setups
	do not cover the heterogeneous data stream case  we face. In our multivariate streaming platform data setting, we show that standard adaptations of statistical procedures do work well in practice.
	
	Ideally, we re-train the $D$ ML models $f_{it}(\cdot)$ each time a new data batch of $B$ demand values becomes available. This would guarantee the best possible forecast performance  for the given ML model based on historical data, and the MLOps engineer does not have to ask the question of when to re-train the model.
	Such an updating scheme is, however, too expensive in terms of labor and computing resources for an on-demand platform to implement in production, as they collect data at high-frequency intervals for a market consisting of many delivery areas (large $D$). 
	To keep the production of accurate streaming forecasts manageable, we therefore suggest  a monitoring scheme that determines when re-training is needed, as detailed next.
	
	\subsection{Machine Learning Monitoring} \label{subsec:monitoring}
	Instead of re-training ML model $f_{it}(\cdot)$ for each data stream at the baseline frequency $t$, 
	we wish to keep the forecast functions mainly constant, as regulated via two  dynamics: One related to the data batch structure and one related to the detection of forecast instability.
	
	First, taking the batch structure of the data into account, we only consider re-training of the ML model whenever a new data batch arrives, so at (end of batch) times denoted by subindices $b : =\{t: \text{mod}(t,B)=0\}$, 
	where mod stands for modulus, being the remainder of dividing $t$ by $B$. Hence, we assume that  the ML model is constant intra-batch: 
	\begin{eqnarray}
		{f}_{ib} =  f_{ib+1} = \ldots = f_{ib+B-1} \quad \forall b. \nonumber
	\end{eqnarray}
	
	Second, at each end of batch time $b$, we perform a stability test to determine whether the ML model should be re-trained or not. 
	We  keep the same ML model (hence no re-training) under ``forecast stability". To define the latter, we introduce additional notation. 
	Denote the forecasts made at batch-end $b$ for data stream $i$ at horizon $q=1,\ldots, Q$ by 
	\begin{eqnarray}
		d_{ib+q|b}^f = \hat{f}_{ib}(\{d_{b+q-j}\}_{j \in \mathcal{J}}, a_{b+q}),    \nonumber
	\end{eqnarray}
	and let
	\begin{eqnarray}
		l_{ib+q}=l(d_{ib+q|b}^f,d_{ib+q}) \nonumber
	\end{eqnarray}
	be the forecast loss, with $l(\cdot,\cdot)$ any loss function computing forecast errors; for instance squared error loss. 
	At the end of the next batch period $b+B$, we compute these forecast losses and investigate whether the forecast loss  for data stream $i$ given by
	\begin{eqnarray}
		L_{ib+B}=\{l_{ib+q}\}_{q=1}^Q  \nonumber
	\end{eqnarray}
	is significantly different, on average, from past losses in a reference batch denoted by $L_{ib}^0$. 
	
	Under the described monitoring  procedure, the null hypothesis of the stability test at time $b+B$ is 
	\begin{eqnarray}
		\mbox{H}_0:  \mbox{E} (L_{ib}^0) = \mbox{E} (L_{ib+B}),  \label{eq:null_hypothesis}
	\end{eqnarray}
	which we test against a two-sided alternative with a standard equality of means test.  In case we reject the null hypothesis for stream $i$,  there is ``forecast instability", the MLOps engineer is alerted to decide on the re-training of the ML model, and  the reference batch is re-initialized.
	In case we do not reject the null, 
	there is ``forecast stability", we keep the same ML model and the losses become part of the reference batch.
	Note that the execution of the standard test requires computing only sample moments and is therefore extremely fast. Therefore, monitoring can be implemented in a production setting for an indefinite amount of iterations. 
	
	The size properties of a standard sequentially implemented test in the case of our non-identically distributed data stream implied loss functions are difficult to assess.\footnote{Recently, research on two sample tests has been mostly focusing on  the high-dimensional iid (i.e.\ independent and identically distributed) data case, e.g., \cite{Zhang_JASA_2020} and \cite{Jiang_JASA_2022}.} 
	However, from our experience, given that the reference batch $L_{ib}^0$ is frequently reset for platform data, and thus also the null, and that the sample size of $L_{ib}$ is  sufficient (i.e.\ $Q=60$ in our application), we expect reasonable size properties of our test. 
	We confirm this with synthetic data in a Monte Carlo simulation study, the results of which are presented in Appendix \ref{app:MonteCarlo}.
	
	\subsection{MLOps Engineer in the Loop} \label{subsec:monitoring_productivity}
	Our monitoring framework  re-trains the  ML models $f_{it}$ for each stream $i$ at different, a priori unknown times, and this through the automatic alert system based on the stability test. To understand the overall forecast system health and take action in case the running ML model needs to be updated or replaced, the MLOps engineer has access to the following metrics at baseline frequency.
	
	First, the time points at which an alert is triggered due to forecast instability that are summarized via the streaming ``re-training" indicator  $r_{it}$ defined as 
	\begin{equation*}
		r_{it} = \begin{cases}
			0 & \quad \text{mod}(t,B) \neq 0 \; \text{OR} \; \left(\text{mod}(t,B)=0 \; \text{AND} \;\text{forecast stability}\right), \\
			1 & \quad \text{mod}(t,B)=0 \; \text{AND} \;\text{forecast instability}.
		\end{cases}
	\end{equation*}
	A  zero value simply indicates a time point of forecast stability of stream $i$, whereas a value of one indicates a time point of forecast instability and hence the necessity for the MLOps engineer to decide on re-training of the ML model. The stream $r_t = \left( r_{1t}, r_{2t}, \ldots, r_{Dt}  \right)$ can thus be used for stability tracking. 
	
	Second,  we compute forecast stability regime durations, which are defined as the number of periods between two successive 1's in $r_t$. This is useful information for the MLOps engineer to directly assess the periodicity of re-training and spot differences among data streams.
	Note that the stream $r_t$ is defined on calendar time and contains historical re-training dates while for future dates it has zero value by default. However, the MLOps engineer can ingest 1's in future dates to activate re-training of the ML model. For digital platforms, this is useful when a region will be impacted by new business contracts that affect the level and dynamics of specific delivery area data streams.
	
	Third, we store the averages of the reference loss and new loss batches, i.e. the empirically counterparts of \eqref{eq:null_hypothesis}, in the stream $P_{it}$ as follows 
	\begin{equation*}
		P_{it} = \begin{cases}
			\left(0 , 0\right) & \quad \text{when}  \;\; r_{it}=0,  \\
			\left( \mbox{Mean} (\widehat{L}_{ib}^0) , \mbox{Mean} (\widehat{L}_{ib+B})\right) & \quad \text{when} \;\; r_{it}=1.
		\end{cases}
	\end{equation*}
	This metric allows tracking the overall accuracy of the ML models. For example, if for a delivery area, the average of the reference loss batch passes a threshold that indicates significant negative business impact, then action is required. Keeping track of both the reference and new loss batch allows gauging the absolute strength of the change.

	\begin{algorithm}[t]
		\caption{\color{black} Monitoring ML Forecast Streams}
		\label{alg:monitor}
		\leading{15pt} 
		\begin{algorithmic}[1]
			\STATE Let $S_{ib}$   be initial data batches for $b=0$ and $i=1,\ldots, D$ where $D$ is the number of data streams
			\STATE Train ML models $ f_{i0} $ on $S_{i0}$, ($i=1,\ldots, D$),  compute initial  forecasts 
			\STATE  Read data batch $S_{i1}$, compute forecast losses to obtain  initial reference batch $L_{i1}^0$
			\FOR{$b = 2, 3, \ldots$}
			
			\FOR{$i = 1$ to $D$}
			\STATE Read data batch $S_{ib}$ 
			\STATE Compute forecast losses $L_{ib+B}=\{l_{ib+q}\}_{q=1}^Q$ 
			\STATE Test for equal average forecast loss with reference batch  $L_{ib}^0$ resulting in $TEST=1 (0)$ if reject (accept)
			\IF{$TEST=1$ \AND MLOps engineer decides to re-train} 
			\STATE Train ML model $ f_{ib}$
			\STATE Reset reference batch $L_{ib}^0$ and set $r_{ib}=1$; $P_{ib}=\left( \mbox{Mean} (\widehat{L}_{ib}^0) , \mbox{Mean} (\widehat{L}_{ib+B})\right)$
			\ELSE
			\STATE Update reference batch $L_{ib}^0$ with $L_{ib+B}$ and set $r_{ib}=0$
			\ENDIF    
			\STATE Compute forecasts $\hat{f}_{ib}(\{d_{b+q-j}\}_{j \in \mathcal{J}}, a_{b+q})$
			\ENDFOR
			\RETURN $D$ forecasts for decision making	
			\ENDFOR
		\end{algorithmic}
	\end{algorithm}
	
	\subsection{Algorithmic View} \label{subsec:algorithm}
	Algorithm \ref{alg:monitor} gives an overview of the  flow of the proposed MLMA monitoring procedure.  Lines 1, 2 and 3 are initialization steps. For each of the $D$ initial data batches, we train the ML models for the first time and compute forecasts. Then we wait until the next batch $S_{i1}$ becomes available, we contrast the initial forecasts with their realizations using the loss function. This results in the first reference batch.
	Lines 4 and 5 run through each incoming data batch and each data stream. Line 6 reads the new available data batch. Line 7 uses this new data batch and the ML model forecasts to compute a new loss batch. Line 8 performs the statistical stability test with null hypothesis given in \eqref{eq:null_hypothesis} and concludes if there is forecast stability or not. Lines 9, 10 and 11 check if there is forecast instability and, if it occurs, re-trains the ML model based on the MLOps engineer's  decision, re-initiates the reference loss batch, and updates the MLOps metric streams. Lines 12 and 13 are active if the statistical test concludes in favor of forecast stability. No re-training of the ML model is required, only the reference batch is expanded with the new forecast losses. Line 15 computes forecasts using the ML model. Finally, line 17 returns the sets of forecasts for all data streams as inputs for operational decision making.

	\section{Monitoring at Digital Platforms} \label{PlatformBusiness}
	
	\subsection{Business Problem} \label{Platform_business_problem}
	Digital platforms are crucial in today’s economy as they connect consumers and businesses, and create substantial economic value by facilitating transactions, enhancing access to information, and optimizing supply chains. On-demand platforms recently generated considerable research interest regarding their economic impact, innovation, and dynamic pricing effects (e.g., \citealp{Burtch_mansci2018, helfat2018dynamic, Chen_JPE_2019, li2024demand, Xu_mansci_2024}).
	
	This paper  exploits the fact that on-demand digital platforms  generate real time data streams about all business aspects. Data stream forecasts are essential inputs for automated decision making at digital platforms. Forecasts determine pricing and logistical aspects such as replenishment of fulfillment centers, or placement of drivers. Given that forecasts are used to actually run the platform, forecast quality has direct business impact. 
	For example in the context of an on-demand delivery platform, forecasting too high demand causes increases in driver's salary, therefore directly increasing the cost per delivery.  Furthermore, platforms have thin profit margins and to be profitable they need to operate at large-scale. 
	
	However, a particular characteristic of digital platforms is that the dynamics of their business changes frequently, due to expansion, competition, changing customer tastes \citep{hu2023fast}. Such a volatile environment is problematic for ML models to maintain the production of accurate forecasts. The problem originates from the fact that monitoring and managing high performance ML forecast models  with existing frameworks is labor intensive. Indeed, such ML models require frequent re-training to deliver high accuracy forecasts, and given the sources of instability, the dates to re-train are unknown a priori. Consequently, the MLOps engineer needs to closely watch each forecast stream in the system and decide if the ML model has to be re-trained. At scale, this would require an entire team of high-skilled, hence costly, professionals.
	
	The MLOps monitoring framework we provide solves this problem and yields strong productivity gains for the MLOps engineer. In fact, much of the monitoring task is offloaded to an algorithm in our MLMA framework. One MLOps engineer can manage the platform's entire forecast system while other engineers in the team can develop new services (e.g.\ go to market forecasting, improvement of ML models with external data). MLMA allows achieving stable forecast accuracies that significantly enhance the operational efficiency of the platform.

	\subsection{Platform Data} \label{PlatformData}
	To illustrate the value of our framework in a real setting, our  analysis involves a unique data set of high-frequency demand streams from a leading on-demand logistics platform in Europe, Stuart, which connects businesses to a fleet of geologically independent couriers.\footnote{The data are provided to us in the context of ongoing research collaborations. Due to a confidentiality agreement, we are not allowed to distribute or report actual demand data. Demand data across all figures are therefore normalized between 0-100. All analyses were carried out with the original data.} 
	We analyze UK London demand data  at the 15-minute frequency from January 1, 2019 
	to March 31, 2021.  London is split into 32  boroughs to efficiently organize parcel deliveries.\footnote{London is split into 32 boroughs plus City of London as administrative area. Since we have no data available for Hammersmith \& Fulham, for ease of exposition, we count 32 boroughs including City of London.}
	All boroughs operate on a daily basis between 9am and 11pm,\footnote{Occasional overnight demand is integrated in the last 15 minutes of the business day.}
	which creates a balanced dataset of 
	$49,320$ 
	observations per borough.
	
	\begin{figure}[ht!]
		\centering
		\includegraphics[width=0.45\textwidth]{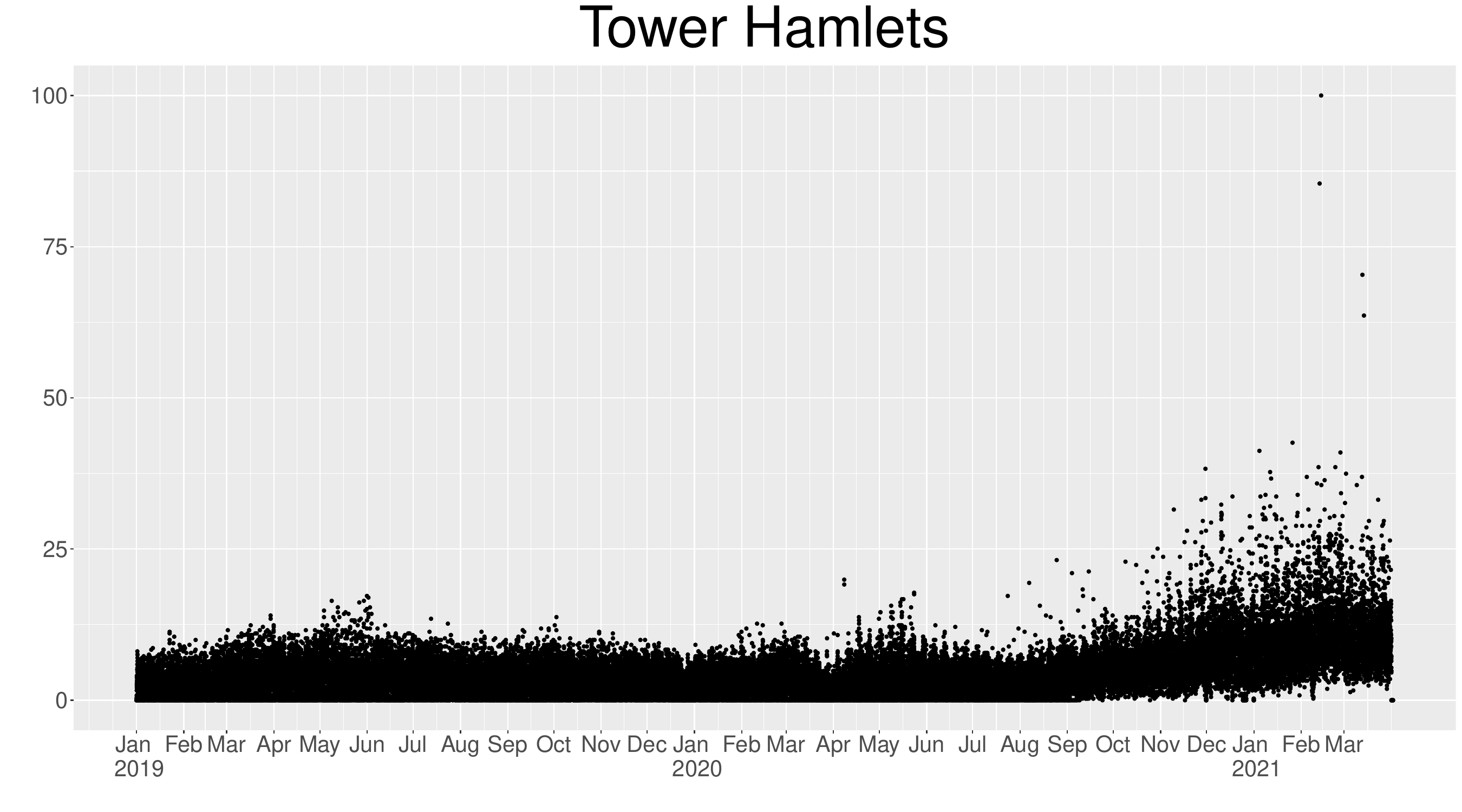}
		\includegraphics[width=0.45\textwidth]{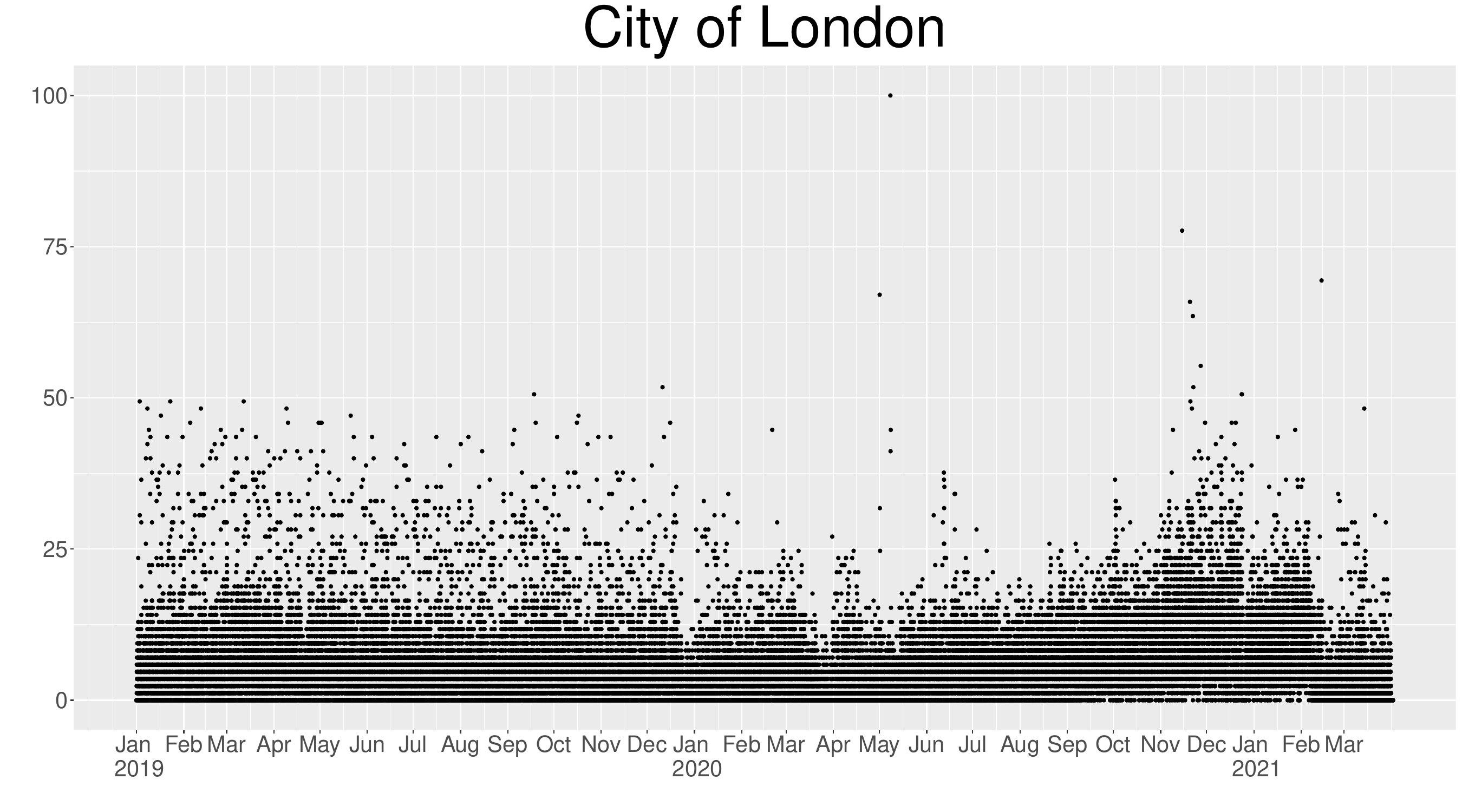}
		
		\includegraphics[width=0.45\textwidth]{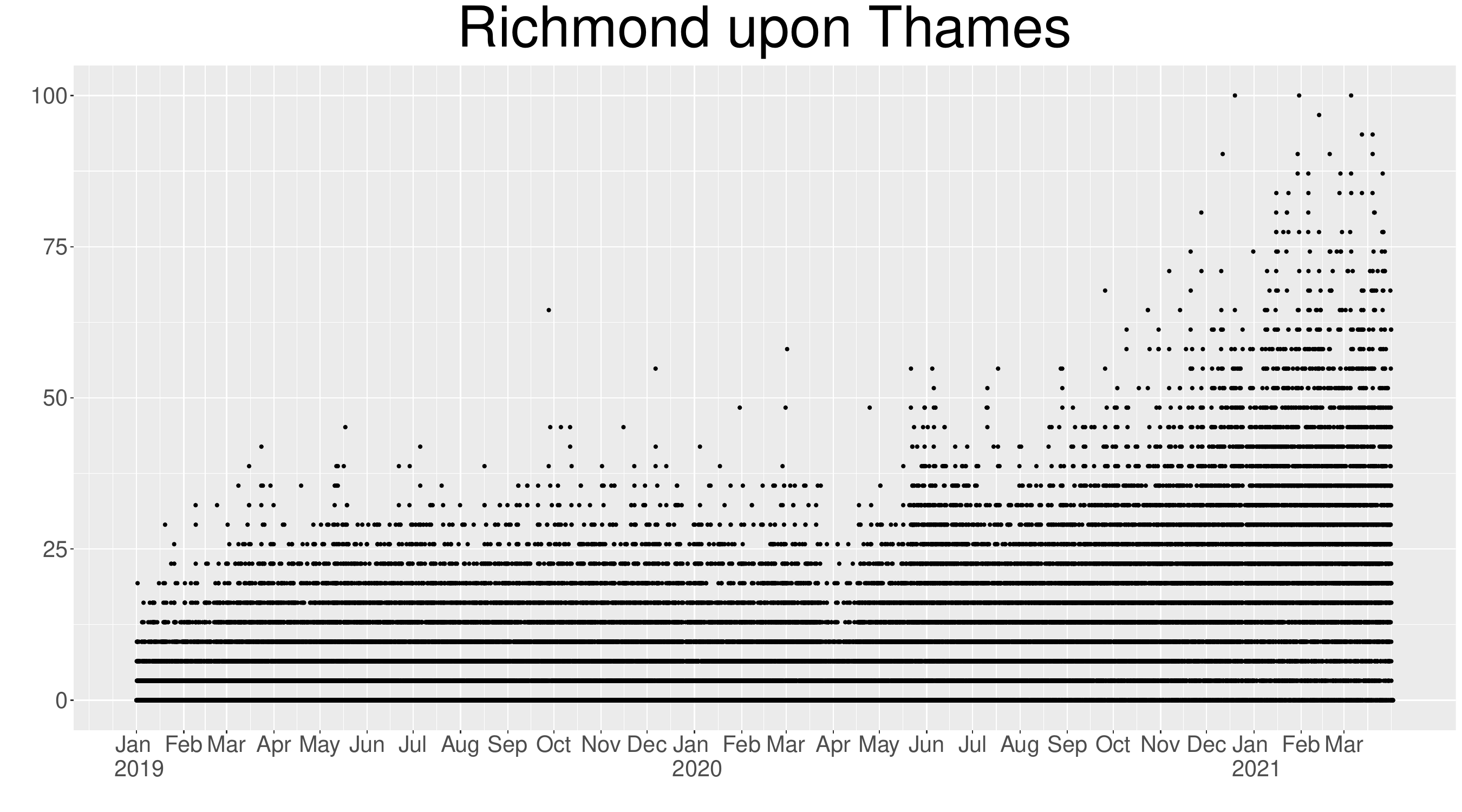}
		\includegraphics[width=0.45\textwidth]{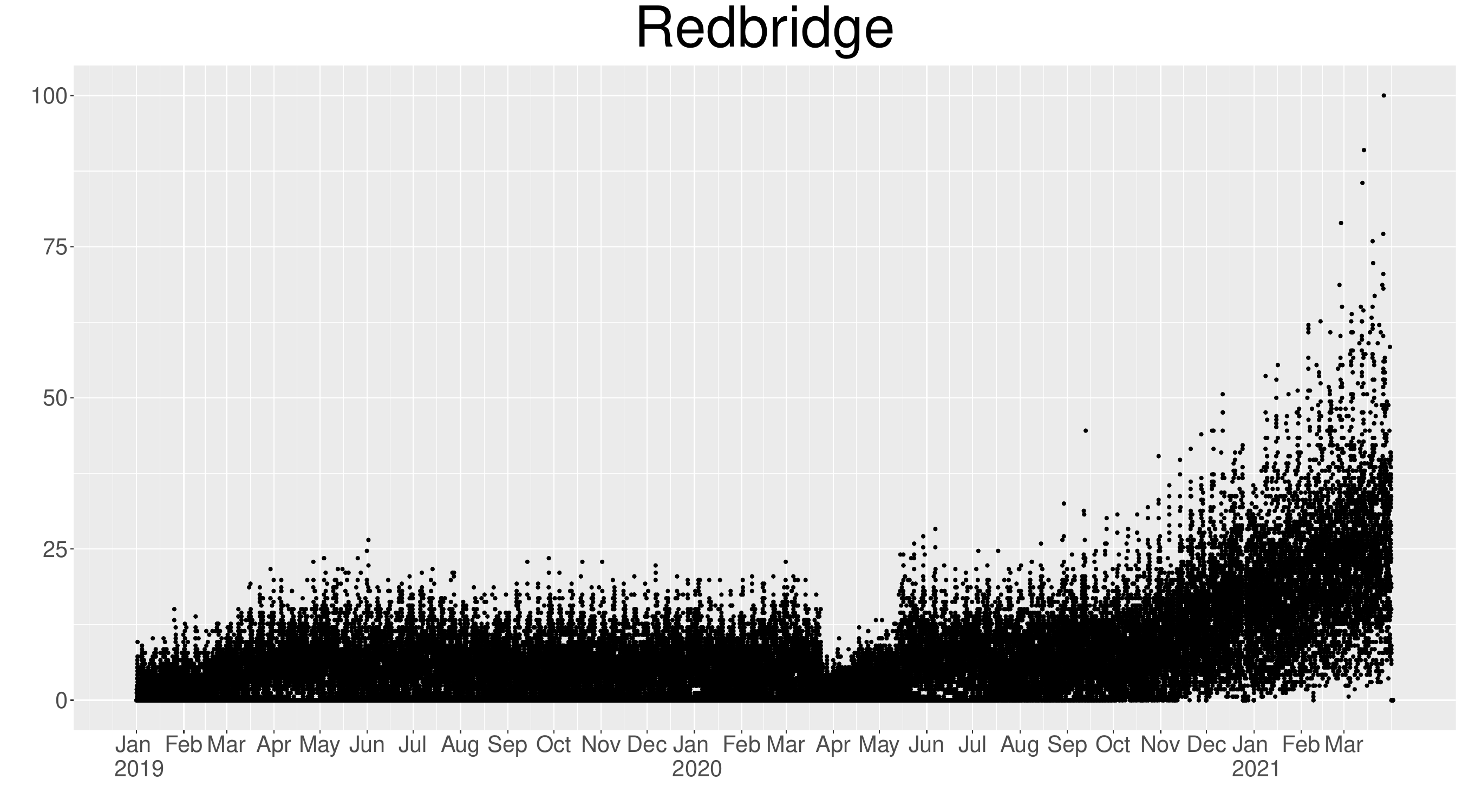}
		
		\includegraphics[width=0.45\textwidth]{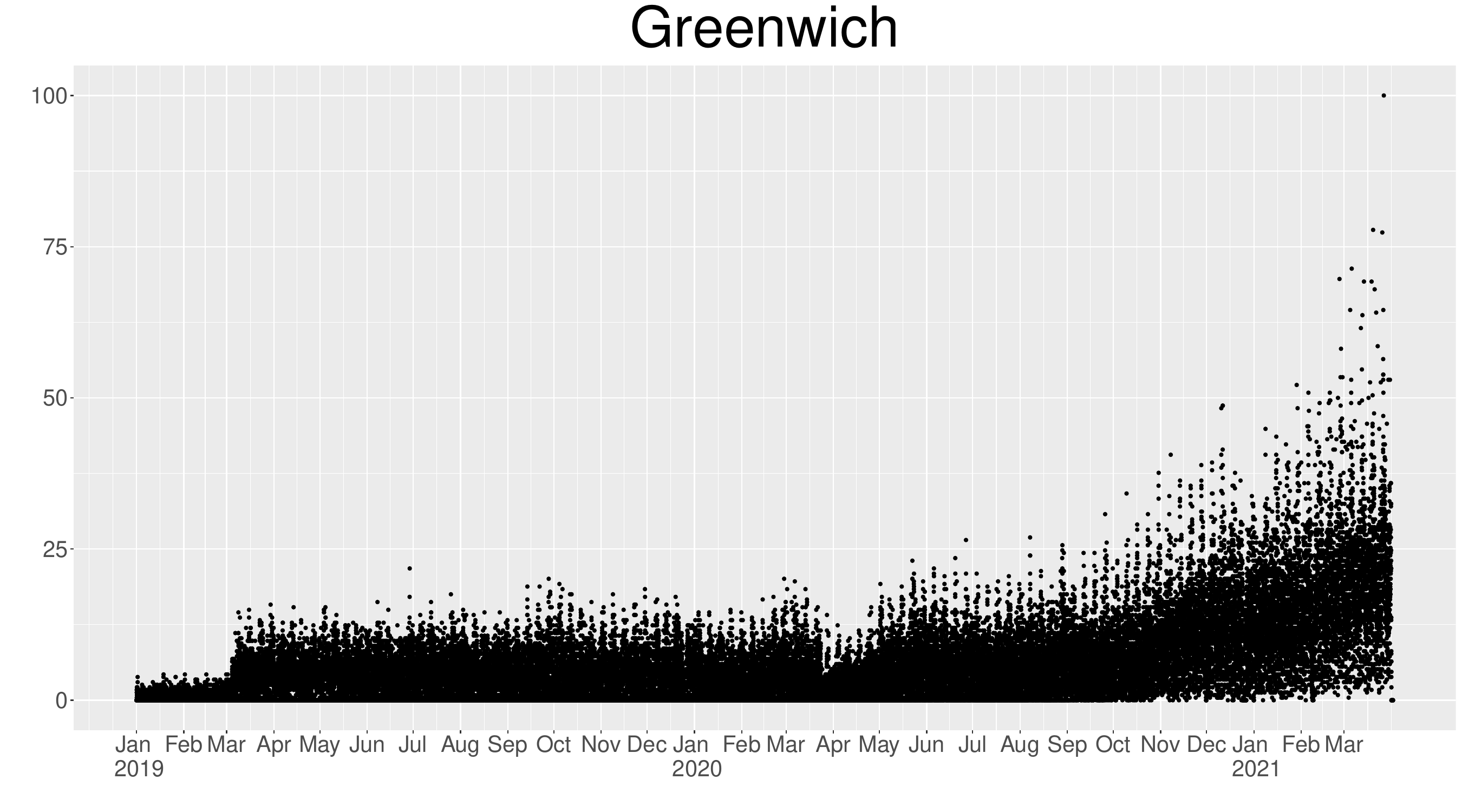}
		\includegraphics[width=0.45\textwidth]{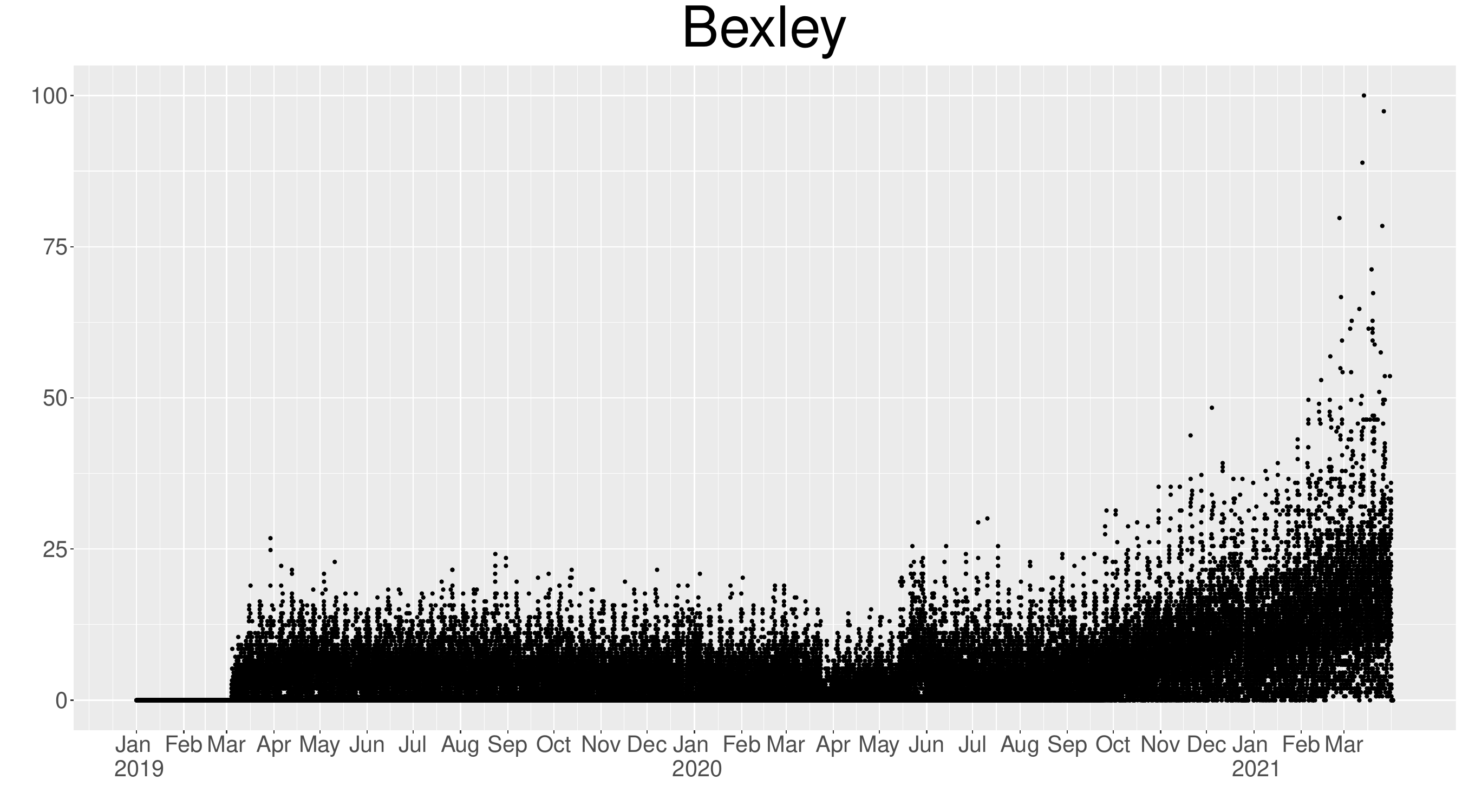}
		\caption{ 15-min demand across six boroughs from January 1, 2019 until March 31, 2021. \label{fig:six_boroughs}}
	\end{figure}

	Figure \ref{fig:six_boroughs} plots  15-minute demand for six representative yet diverse boroughs. 
	In Tower Hamlets, the borough with the highest total demand, demand is stable ranging between (roughly) zero and ten until early 2021 when minimum demand starts increasing and becomes more volatile, thereby mounting to its highest values in the sample. 
	City of London is characterized by high intra-day peak demand over the entire period requiring careful fleet planning throughout the day. 
	In contrast, Richmond upon Thames has demand characterized by regular demand at specific hours. 
	
	Although visually quite different, two common distinctive features of this streaming demand data compared to more traditional time series appear. 
	First,  demand is unstable. Some boroughs have low initial demand before jumping to higher levels.
	The Covid-19 pandemic and first March 2020 lockdown result in a demand drop for several weeks (see Redbridge), but then demand quickly starts to grow rapidly with higher volatility until the end of the sample. 
	Second, the overall range in demand is high. For example, Greenwich and Bexley start in 2019 with low variation close to zero demand and end with demand fluctuating between 0 and 100. 
	This nonstationarity due to drifting levels and sudden variation of demand data streams necessitates an MLOps monitoring framework that can trigger the re-training of ML models. Given the large scale of the problem, many streams  measured at high frequency, this re-training can only be done following an automated decision process with the MLOps engineer in the loop via key metrics; all features that MLMA offers.

	\begin{figure}
		\centering
		\includegraphics[width=0.8\textwidth]{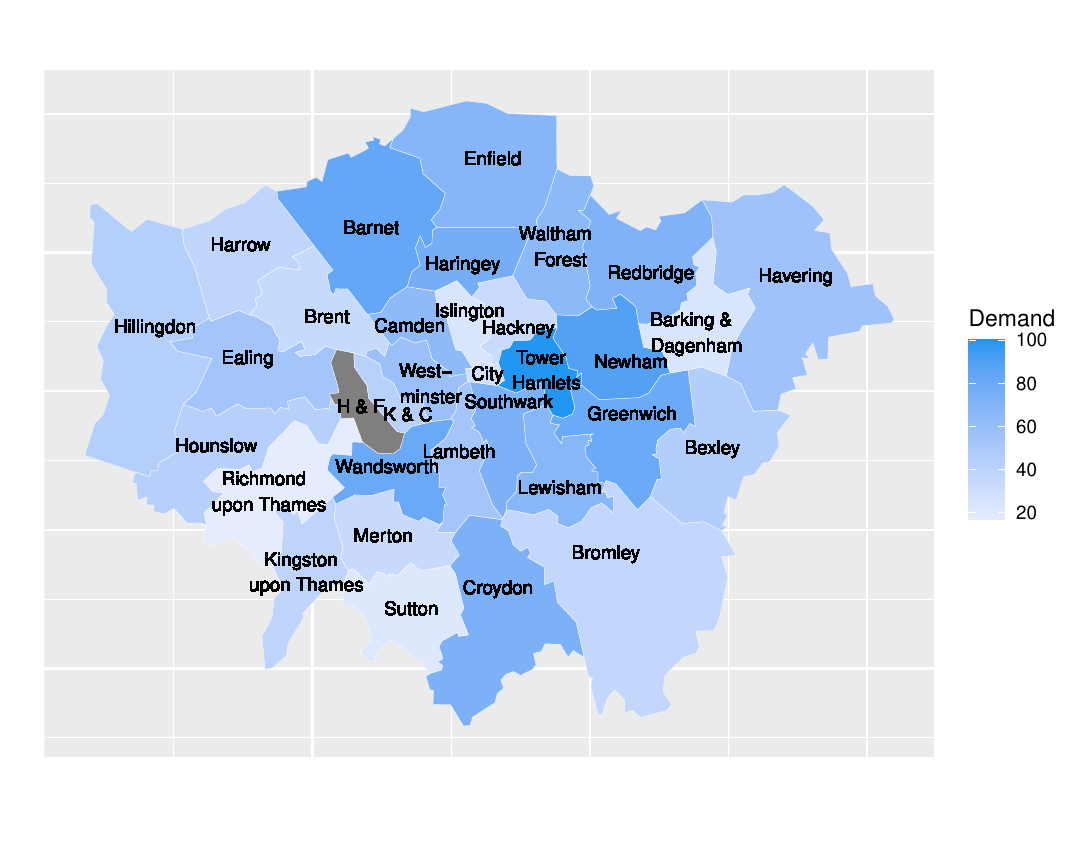} \vspace{-1.5cm}
		\includegraphics[width=0.8\textwidth]{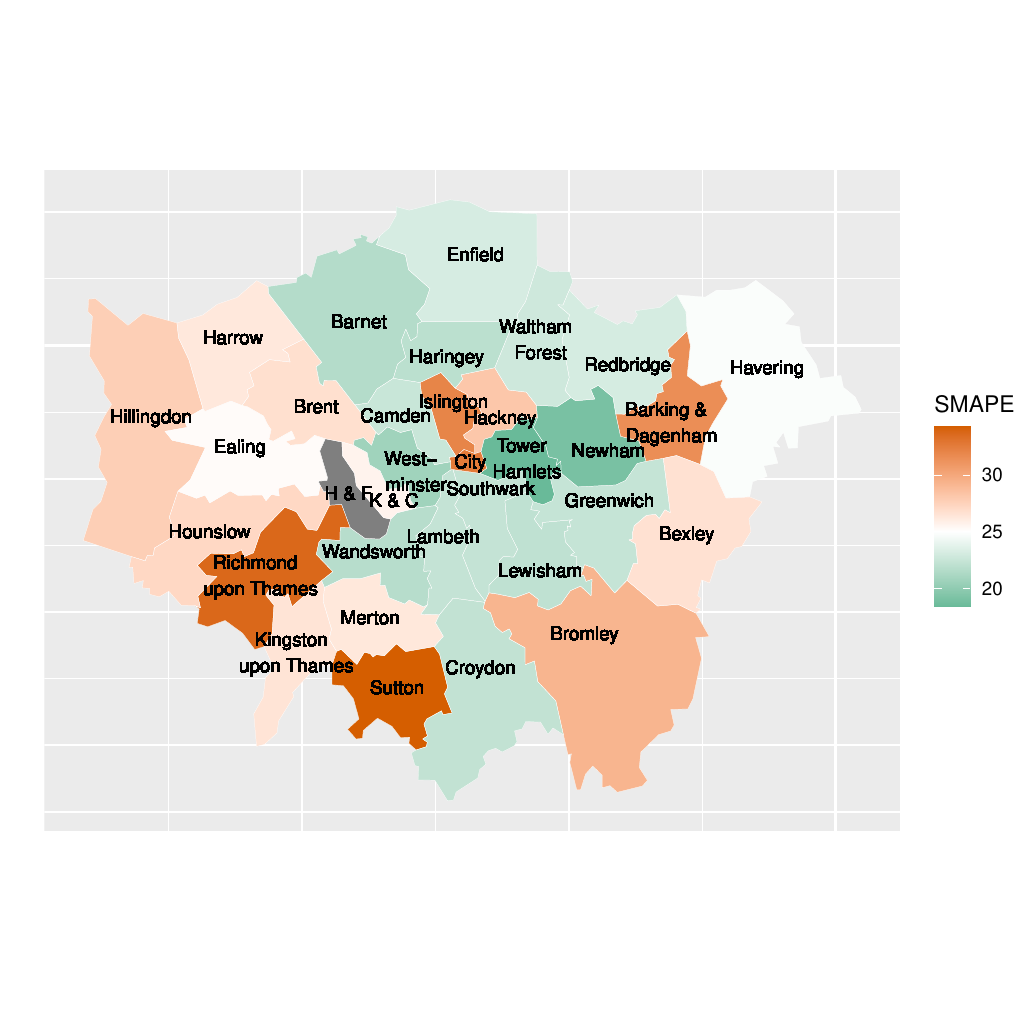} \vspace{-1cm}
		\caption{Heat maps of demand in each borough (top: the darker the shading, the higher the demand) and forecast  errors (bottom: above, below and at median error in respectively red, green and white) for MLMA, to be discussed in Section \ref{sec:forecast_results}.} \label{fig:heat_maps_london}
	\end{figure}
	
	Figure \ref{fig:heat_maps_london} 
	(top) visualizes total demand in our sample on a London map to highlight the geographical demand differences across boroughs.\footnote{Hammersmith and Fulham (H \& F) is colored gray since no demand data is available for this borough.} Dark (light)  blue boroughs are characterized by high (low) demand. 
	Demand is not uniformly distributed; the main reason for this is that fleet planning is organized with respect to drivers' availability rather than clients' demand in order to optimize the drivers' compensation schemes. This  gives rise to substantial borough-demand heterogeneity ranging from residential over business towards tourist areas. Additional insights into the seasonality patterns of the platform data streams and the time-varying dynamics between the boroughs are described in Appendix \ref{app:figures}.
	Successful ML models have to be flexible enough to capture these specific features of our demand data. However, accurate ML models have to be re-trained when the underlying connections with the target variable break down. The fact that this happens at a priori unknown times calls for our monitoring framework.

	\subsection{Monitoring Set-up} \label{MLMA_setup}
	We  discuss the monitoring set-up we use to compare the performance of our MLMA framework against other monitoring benchmarks.
	
	\subsubsection{ML Model} \label{subsec:treeML}\
	The key ingredients in the initial setting of the MLMA framework concern the choice of ML model $f_{it}(\cdot)$ and predictors in equation \eqref{eq:generalmodel}.
	While our MLMA framework allows one to use their ML model of preference, we use random forest \citep{Breiman2001} as ML model as it has been carefully selected for usage on our platform data and  therefore is the preferred ML model as part of the initial setting of our MLMA framework in Figure \ref{fig:MLOps_framework}.\footnote{The choice for a tree-based ML model over a neural network or deep learning model is a deliberate one, and motivated by their routine availability of robust, sophisticated and sensible ``off-the-shelf' implementations across standard software (e.g., \citealp{januschowski2022forecasting}). 
		When time is short as for digital platforms, the ability to rely on a mature, readily available model implementation is key. While several tree-based ML-models have been explored, random forest generated the most accurate and stable forecasts, therefore explaining its choice as ML model in the initial setting of our MLMA framework.  Similar insights for MLMA compared to its benchmarks were obtained when using XGBoost-- the runner-up --instead of random forest, which underwrites the robustness of MLMA to the choice of ML model. Results are available upon request.
	}
	
	A random forest estimates $f_{it}(\cdot)$ by combining regression trees. A regression tree is a nonparametric method  that partitions the feature space to compute local averages as forecasts, see \cite{Efron_Hastie_2016} for textbook explanations. The random forest hyperparameters are the number of trees that are used in the forecast combination, the number of features to randomly select when constructing each regression tree split, and the minimum number of observations in each terminal node to compute the local forecasts. 
	Since hyperparameter tuning is not the main objective of our paper and default settings often attain excellent performance across various settings, 
	we use the standard implementation of the \texttt{randomForest} package \citep{randomForest} in \verb|R| \citep{Rcoreteam}.
	
	The features in the random forest model \eqref{eq:generalmodel} are selected based on domain expertise: We include  a linear trend, hour-of-the-day and day-of-the-week seasonal dummies in $a_t$; totaling 32=1+15+6 features in $a_t$ of model \eqref{eq:generalmodel}.
	For the lag structure of the $D$-dimensional historical data stream $\{d_{t-j}\}_{j \in \mathcal{J}}$, we take
	$\mathcal{J} = \{60, 420\}$, implying ``same quarter hour last day" $(d_{t-60})$ and ``same quarter hour last week" $(d_{t-420})$ lags for all of the $D=32$ boroughs; totaling $64=2\times32$ features in $d_{t-j}$ of model \eqref{eq:generalmodel}. By including lags of all the $D$ boroughs, we account for  spillovers across boroughs. 
	As a result, we use a random forest model with a total of $p=86$ predictors to forecast demand.
	From extensive experimentation with the platform demand data, this ML model was found to pick up the most relevant information while keeping the dimension reasonable.
	
	\subsubsection{MLMA Framework}\label{subsec:forecast}
	We implement the following monitoring forecast procedure for each of the 32 boroughs in London. 
	To ensure data availability across all boroughs, we use data as of March 1, 2019 and start training  each ML model on an initial sample period of 180 days of 15-minute data. At business day-end, we produce forecasts for quarter hour demand  $d_{ib+q|b}^f$ (from 9am to 11pm) for the next day, i.e.\ a forecast horizon $q=1,\ldots, B=60$.\footnote{Since we forecast all quarter hours for
		the next day, the forecast horizon $Q$ and batch size $B$ are equal (namely 60) in our application. To facilitate readability, we simply use $B$ to denote both in this section.} 
	
	The out-of-sample forecast monitoring evaluation period  runs from August 28, 2019 to March 31, 2021. 
	This generates a daily stream of quarter hour forecasts that are evaluated against  actual demand  to obtain the average daily loss batch $\frac{1}{B} \sum_{q=1}^B (d_{ib+q} - d_{ib+q|b}^f)^2$. 
	The stability test investigates if this batch is significantly different from the reference batch.
	To implement the standard equality of means stability test, we use  the function \texttt{t.test} (with default arguments) in \verb|R| and set the size of the test at five percent. 
	
	If the stability test does not reject, then forecasts are computed with the existing forecast function. 
	If the stability test rejects, the MLOps engineer receives a re-training alert, and based on the metrics decides to  re-train the ML model or not. Based on industry discussions with digital platform domain experts,  the MLOps engineer decides to re-train when the average loss exceeds two standard deviations of the symmetric mean absolute percentage error (mathematically defined in the next paragraph) computed on the initial sample. In addition, the MLOps engineer decides to pro-actively re-train the ML model on certain event dates that are expected to impact the demand dynamics in the London area.\footnote{The event dates, set based on industry discussions, consist of all bank holidays, the start of the major school holidays in the London area and Wimbledon as major sport event. \label{footnote:events} }  Re-training is always done on the most recent 180 days (hence with the same sample size) in order to keep the computation time constant. To appreciate the efficiency of MLMA in terms of human time spent,  we also implement a manual version, labeled ``Manual MLMA" where the MLOps engineer actually executes our framework by hand.

	To evaluate  out-of-sample forecast accuracy, we use the symmetric absolute percentage error loss  $L^{\mbox{sape}} (d_{ib+q},d_{ib+q|b}^f)= 100 \mid d_{ib+q} - d_{ib+q|b}^f \mid / (|d_{ib+q}| + |d_{ib+q|b}^f|)$,  a popular  metric at delivery platforms because of its
	(i)  in-built asymmetry where under-forecasting (forecast below actual) is penalized more than over-forecasting (forecast above actual) and (ii) relative nature which facilitates comparisons across heterogeneous boroughs. We average $L^{\mbox{sape}} (d_{ib+q},d_{ib+q|b}^f)$ across the out-of-sample quarter hours and denote this as the SMAPE.

	\subsubsection{Monitoring Benchmark Strategies}\label{subsec:alternative_tests}
	We compare the performance of our MLMA framework with the three benchmark monitoring strategies: (i) do nothing, (ii) periodic re-training, and (iii) on demand monitoring.
	
	The first ``Do nothing" strategy never re-trains the ML model. We consider two choices
	under this strategy. As a first choice, labeled as ``Do nothing: ML", we estimate the random forest model once on the initial data batch, and use the corresponding forecast function throughout the whole evaluation period. That is, we train a high quality ML model once and use new incoming batches only to produce forecasts.
	As a second choice in the no-monitoring strategy, we use a very simple model which is not determined by the initial data batch. 
	To this end, we     take $d_{ib+q|b}^f =  d_{ib+q-420}$ which we call the ``Do nothing: Naive" strategy since  it simply implies that the forecast of a specific quarter hour is the value observed in the same quarter hour and day of the  previous week. This way of forecasting is popular in the industry since it is ultra fast, i.e.\ it does not require ML model building nor parameter estimation, and it works well in the case of strong seasonality patterns as observed in platform  data. In addition, the MLOps engineer has no monitoring task. However, while appealing a strategy in terms of used resources, the use of actively monitored  advanced ML models typically allows for much better forecast accuracy, resulting ultimately in better business performance of the platform.
	
	The second benchmark monitoring strategy consists of  
	re-training the ML models for all boroughs at fixed periodicity, which is in contrast to our monitoring approach that triggers re-training at prior unknown times that may vary across boroughs. 
	We re-train the ML models both daily (``Periodic: Daily") and every semester (``Periodic: Semester") for each borough.
	Daily automatic re-training
	serves as an empirical performance upper bound, but is computationally unfeasible in the streaming data environment for which we develop our monitoring approach. Nevertheless, despite the one-time heavy computational cost to report results in this paper, such a  benchmark allows contrasting how much forecast performance is lost by re-training only in case of significant instability.
	Semester-based automatic re-training is  easy to implement, though it  requires large computational resources to re-train all ML models simultaneously, and the MLOps engineer collects no forecast instability information.   
	
	The third benchmark strategy is on demand monitoring. In our platform application, this demand originates from the business teams.
	Re-training should occur when the SMAPE forecast loss of the reference batch goes above a certain business-set threshold. Once the threshold is exceeded, there are significantly higher operating costs that squeeze borough specific profit margins, hence re-training should be triggered to prevent this. Based on industry discussions, we set this threshold at 30\% and label the  monitoring strategy ``On demand". 
	
	\section{Results} \label{sec:forecast_results}
	We discuss the monitoring performance of MLMA and the considered benchmarks in terms of forecast accuracy (Table \ref{tab:SMAPE_all_boroughs}), labor costs (Table \ref{tab:Labour_all_boroughs}), and computing costs (Table \ref{tab:computing_time_all_boroughs}). 
	
	\begin{table}[t]
		\caption{SMAPE forecast performance, averaged over all boroughs, of MLMA and the benchmark monitoring methods. The bottom row displays the increase or decrease in SMAPE of the benchmarks relative to MLMA.} 
		\label{tab:SMAPE_all_boroughs}
		\resizebox{\textwidth}{!}{\begin{minipage}{\textwidth}
				\centering
				\begin{tabular}{lcccccccccccc}
					\hline
					& MLMA &&&   \multicolumn{2}{c}{\underline{Do nothing}} && \multicolumn{2}{c}{\underline{Periodic}} && \multicolumn{1}{c}{\underline{On demand}}   \\
					&  &&& ML  & Naive  &&  Semester &   Daily &&  \\
					\hline
					& 25.36 &&& 42.38 & 28.28 && 30.48 & 22.96 && 28.63   \\
					&  &&& +67.11\% &  +11.51\% && +20.19\% & -9.46\% && +12.89\%  \\ 
					\hline
				\end{tabular}
		\end{minipage} }
	\end{table}
	
	Table \ref{tab:SMAPE_all_boroughs} 
	reports average forecast accuracy measured in SMAPEs for all London boroughs across the monitoring strategies. Details for each specific borough are reported  in Table \ref{tab:SMAPE-per-borough} of Appendix \ref{app:figures}.  Overall, MLMA yields an average SMAPE of
	25.36. 
	At the disaggregated level (see Table \ref{tab:SMAPE-per-borough}  or Figure  \ref{fig:heat_maps_london}, bottom panel), boroughs such as City of London and Sutton are more difficult to forecast and have average SMAPEs slightly above 30. 
	In contrast, some boroughs, Tower Hamlets and Newham, reach average SMAPEs below 20. While the MLMA  SMAPEs are interesting on their own because they measure the ML model performance, the main purpose is to compare MLMA average levels of SMAPE with those of the benchmark monitoring strategies. 
	
	We start with the two ``Do nothing" strategies. Training the ML model on the initial data batch only and keeping the model fixed afterwards (``Do nothing: ML") implies an average SMAPE of 42.38 across boroughs. This is 
	67\% higher than MLMA. This sizable gain occurs due to the strong instability of the ML models for platform data streams. Re-training is thus a necessity to keep  accuracy high. The second ``Do nothing: Naive" strategy 
	performs in contrast much better with an average SMAPE of 28.28, or 
	12\% higher than MLMA.
	We next turn to the periodic monitoring strategies. The six month periodic re-training strategy obtains an average SMAPE of 30.48 across boroughs, which is
	20\% higher than MLMA. The computationally unfeasible daily re-training exercise results in an average SMAPE of 22.96, and constitutes overall the smallest SMAPE as expected. Compared to MLMA, this is only 
	9\% lower which implies that  MLMA achieves excellent accuracy by re-training much less frequently (as discussed below). Finally, the on-demand strategy attains an average SMAPE of 28.63, which is 13\% higher than MLMA.
	
	\begin{table}[t]
		\caption{Labor costs (average number of hours spent and salary in USD per month), totaled over all boroughs, of MLMA and the benchmark monitoring methods. The bottom row displays the increase or decrease in labor costs of the benchmarks relative to MLMA.} 
		\label{tab:Labour_all_boroughs}
		\resizebox{0.95\textwidth}{!}{\begin{minipage}{\textwidth}
				\centering
				\begin{tabular}{llcccccccccccc}
					\hline
					&  MLMA & Manual &&   \multicolumn{2}{c}{\underline{Do nothing}} && \multicolumn{2}{c}{\underline{Periodic}} && \multicolumn{1}{c}{\underline{On demand}}   \\
					&    & MLMA && ML  & Naive  &&  Semester &   Daily &&   \\
					\hline
					Hours &  11.97 & 59.97&& 0.08 & 0.00 && 0.25 & 49.01 && 13.22    \\[0.2cm]
					\hline
					Salary &  775.06 & 3,883.06 && 5.18 & 0.00 && 16.19 & 3,173.40 && 855.99   \\[0.3cm]
					& & +401.00\% && -99.33\% &  -100.00\% && -97.91\% & +309.44\% && +10.44\% \\ 
					\hline
				\end{tabular}
		\end{minipage} }
	\end{table}
	
	Table \ref{tab:Labour_all_boroughs} shows the implied MLOps engineer labor costs for all boroughs together, measured in average number of hours spent per month, and implied salary cost for each of the monitoring strategies. Based on discussions with MLOps engineers from the industry, it would take about one minute to validate a re-training alert. In addition, re-training an ML model also requires passing a test protocol. For a mature platform company, this protocol requires little intervention from a data engineer and takes about 3 minutes. From Glassdoor, we find 124,316 USD as an average yearly salary for a London based senior MLOps engineer which implies 64.75 USD per hour based on a 40 hours work week.  
	Based on this information and the number of observed re-training alerts, the MLMA framework costs about 12 hours per month or on average about 22 minutes  per borough. Expressed in salary cost, this implies respectively 775 and 24 USD. This monthly cost can vary considerably among boroughs. For example, the relatively stable Tower Hamlets borough has an average labor cost of 15 minutes while the unstable Bexley borough costs almost twice (32 minutes). 
	
	To put the MLMA labor cost into perspective, we compare it  with manually implementing the testing approach at the core of the MLMA framework (``Manual MLMA"). More specifically, we measure the time it takes for an MLOps engineer to perform the statistical tests on a daily basis. Such a task takes about four minutes per day for one borough, that is three minutes for the test and one minute for deciding to re-train according to the SMAPE check if the test rejects. This manual way of monitoring implies a cost of 60 hours (3,883 USD) per month, or a 400\% increase compared to MLMA. This is more than a week per month which explains why companies are often reluctant to bring advanced ML models into production. The ``Do nothing"   monitoring strategies have negligible labor costs as they have no human in the loop.  The periodic re-training strategy at each semester also has negligible labor costs. However the daily re-training scheme, while eliminating the decision to re-train or not, has a large labor cost coming from the test protocols. In fact,  daily re-training implies a monthly cost of 49 hours  (3,173 USD), or about 300\% higher than MLMA. On-demand monitoring   has slightly higher labor cost than MLMA (+10\%). 
	
	\begin{table}[t]
		\caption{Computing costs expressed in CPU and CO2 (in USD per year), totaled over all boroughs, of MLMA and the benchmark monitoring methods. The bottom row displays the increase or decrease in CO2 cost of the benchmarks relative to MLMA. } 
		\label{tab:computing_time_all_boroughs}
		\resizebox{\textwidth}{!}{\begin{minipage}{\textwidth}
				\centering
				\begin{tabular}{lccccccccccccc}
					\hline
					&& MLMA &&&   \multicolumn{2}{c}{\underline{Do nothing}} && \multicolumn{2}{c}{\underline{Periodic}} && \multicolumn{1}{c}{\underline{On demand}}   \\
					&&  &&& ML  & Naive  &&  Semester &   Daily &&  \\
					\hline
					CPU  && 207.36 &&& 7.20 & 0.44 && 5.28 & 1114.32 && 230.40  \\[0.2cm]
					\hline
					C02  && 2.55 &&& 0.09 & 0.07 && 0.14 & 13.72 && 2.84  \\[0.3cm]
					&&  &&& -96.53\% &  -99.79\% && -97.45\% & +437.38\% && +11.11\%  \\ 
					\hline
				\end{tabular}
		\end{minipage} }
	\end{table}

	Table \ref{tab:computing_time_all_boroughs} collects the average yearly monitoring computing costs for the 32 boroughs. Each time a model is re-trained there is a CPU usage cost that we measure in hours. In addition, there is a daily small (one second) CPU usage cost to compute the forecasts. Note that the compute time heavily depends on the complexity of the used ML model, and therefore what matters most is the relative differences between the different monitoring strategies. In our case of a random forest model, training time is about 3.5 minutes and is relatively fast compared to deep learning models.  We report two metrics for computing cost which are directly related to compute time. First, the CPU cost for renting a virtual machine (m5.4xlarge) on AWS which cost at the time of writing 1.624 USD per hour. 
	Second, the CO2 offset cost of the compute time. This metric is increasingly important for platform businesses that typically have strategically important carbon footprint goals. To transform computing time in hours to CO2 offset cost in USD, we proceed as follows. In terms of hardware, we consider a CPU chip consuming 1k Watt per hour. In the UK, in our sample period the CO2 content per Mega Watt hour is 232 kilogram, and the cost of 1000 kilogram of CO2 is 86.25 USD. Therefore the cost for 1k Watt hour is 0.02 USD ($232\times86.25/1000^2$). 
	MLMA  costs yearly 207.36 and 2.55 USD in terms of CPU and CO2 metrics respectively. Compared to the ``Do nothing" and periodically re-training each semester strategies, this is much higher than expected because the ML model is  never or hardly re-trained. In contrast, deterministic re-training at daily frequency implies a cost of 1,114 and 13.72 USD per year for the CPU and CO2 metrics respectively, amounting to a 437\% increase compared to MLMA. The on-demand strategy has a slightly higher cost than MLMA. 
	Finally,  while the yearly computing costs might appear low at first sight, note that these costs are reported only for the 32 boroughs constituting the London market in the setting of our case study. However, the number of data streams to handle for  platform businesses is much higher (easily thousands) and the costs linearly increase with this number.

	\section{Implications}\label{sec:implications}
	This section consists of two parts. In Section \ref{subsec:results:implications}, we focus on business implications and discuss the link between forecast accuracy and total costs, and their geographical differences. In Section \ref{subsec:results:kpis}, we provide insights about our MLMA framework in terms of the metrics reported to the MLOps engineer. 
	
	\subsection{Balancing Forecast Accuracy and Resource Costs} \label{subsec:results:implications}

	\begin{figure} 
		\centering
		\caption{Trade-off between forecast performance (SMAPE) and costs (human and computational resources) for MLMA and the other monitoring strategies across all boroughs (left); and MLMA for the 32 boroughs (right). 
		}
		\includegraphics[width = 0.45\textwidth]{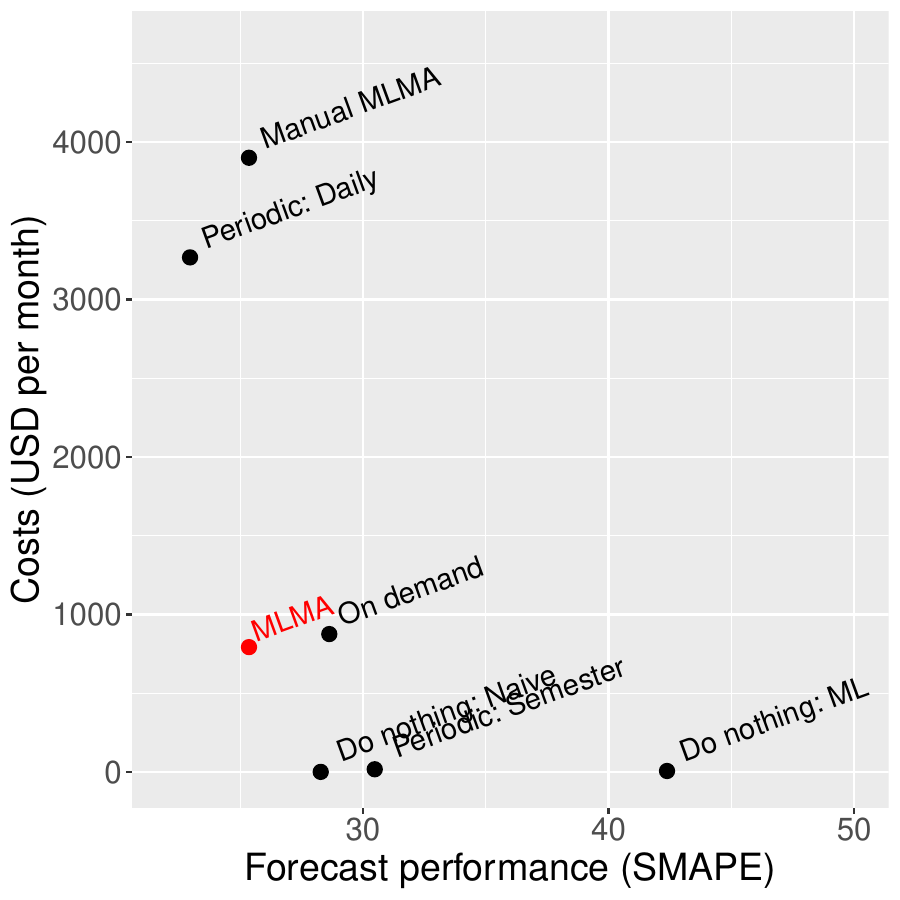}
		\includegraphics[width = 0.45\textwidth]{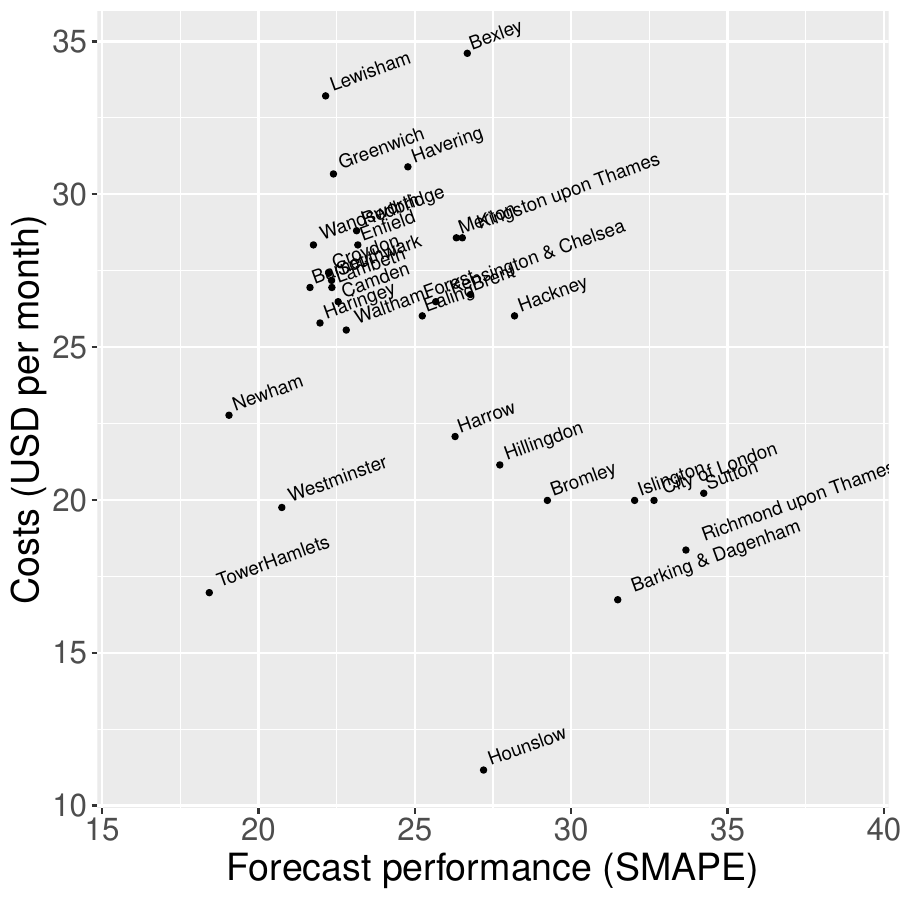}
		\label{fig:scatter_smape_cost}
	\end{figure}

	MLMA  performs well in terms of forecast accuracy, but costs time in human and computational resources. Overall, how does MLMA compare to the alternative monitoring strategies? Figure \ref{fig:scatter_smape_cost} (left) plots forecast accuracy (from Table \ref{tab:SMAPE_all_boroughs}) against 
	total monthly costs (in terms of labor and computational resources; the total of Tables \ref{tab:Labour_all_boroughs} and \ref{tab:computing_time_all_boroughs}) of the different strategies. Low-cost strategies always pay a price in forecast accuracy compared to MLMA, and in addition can be risky because they keep the MLOps engineer almost entirely out of the loop. The daily re-training strategy is more accurate than MLMA but is accompanied with costs too high to put in production in platform businesses typically need to optimize their operations on thousands of data streams. Faced with this trade-off between accuracy and costs, MLMA strikes a good balance. In fact, the desired forecast precision can be attained with ML models in production without incurring excessively large labor and computational costs (less than 1,000 USD per month for the 32 boroughs, or roughly 30 USD per borough). The MLOps engineer is sufficiently kept in the loop thanks to automated statistically test based alerts and metrics (see Section \ref{subsec:results:kpis}). The manual variant of MLMA yields the same forecast precision, but takes too much time from the MLOps engineer (i.e.\ total monthly cost around 4,000 USD). 
	
	Figure \ref{fig:scatter_smape_cost} (right) sheds a more granular light on the forecast accuracy versus cost relationship by considering only MLMA for all boroughs separately. In fact, this is a zoom in the red MLMA dot of the left panel of Figure \ref{fig:scatter_smape_cost}. A small cluster of boroughs for which demand is more complex to forecast, and have therefore relatively high SMAPEs, are experiencing lower monitoring costs. This relationship does not hold for all boroughs though, and as a matter of fact the most accurate borough Tower Hamlets is also the cheapest. Therefore, increase costs does not necessarily translate in better forecast performance. The latter can also been seen from the second cluster of boroughs that have total costs between 25 and 35 USD for roughly the same SMAPE levels.
	
	Finally, we investigate the existence of a relationship between forecast accuracy and geographical location.  This can be seen in Figure \ref{fig:heat_maps_london} (bottom panel) which displays the SMAPE London heatmap for the random forest monitoring procedure. SMAPEs range  between 18 (Tower Hamlets) and 33 (Sutton). There is no specific dependence between geographic location and forecast performance. However, comparing with Figure \ref{fig:heat_maps_london} (top panel), we identify a strong negative relationship between demand volume and SMAPE, with a linear correlation coefficient of -0.92. This is important information when the platform business optimizes the size of its delivery zones, keeping into account that too large zones are inefficient for managing and compensating couriers, i.e.\ the supply side of the platform.

	\subsection{MLOps Engineer Metrics Insights} \label{subsec:results:kpis}
	The MLMA framework generates metrics that allow MLOps engineers to make 
	re-training decisions in case of forecast accuracy  instability alerts. 
	MLMA allows computationally efficient re-training of ML models at dates where forecast loss streams indicate instability. These dates are unknown a priori, which is in contrast to the strategy of re-training at a pre-specified frequency. We now report details on the frequency and timing of re-training the random forest ML model.

	\begin{figure}[t]
		\centering
		\includegraphics[width=0.75\textwidth]{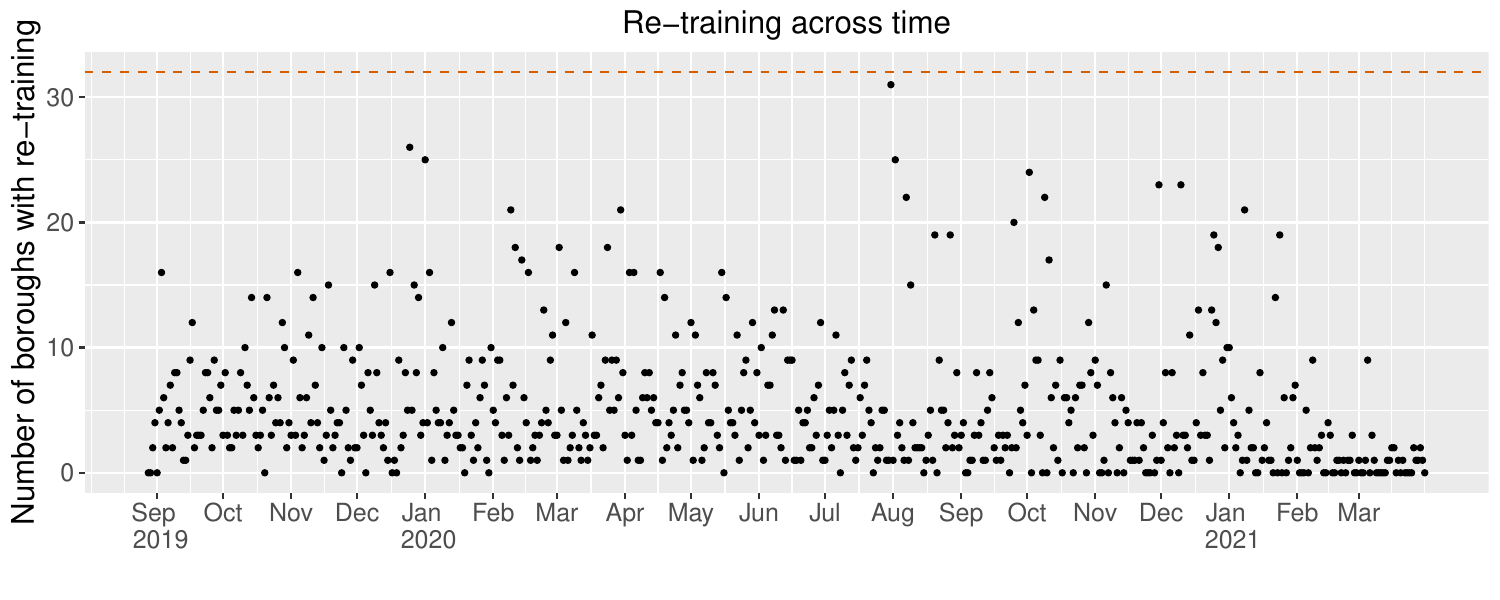}
		\includegraphics[width=0.75\textwidth]{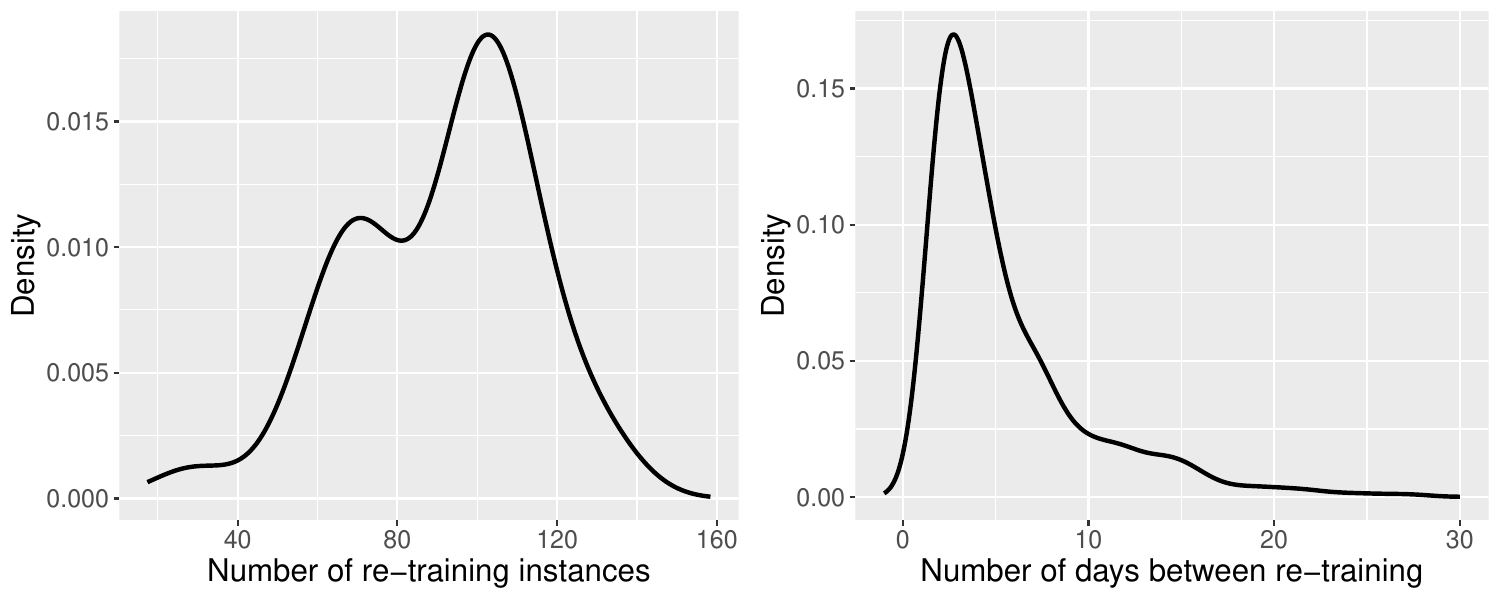}
		\caption{Number of re-training instances  for MLMA across time (top) together with density estimates for the number of re-training instances and number of days between re-training dates (bottom).
			\label{fig:break_information}}
	\end{figure}

	Figure \ref{fig:break_information} (top panel) displays for each day in the out-of-sample evaluation period the number of boroughs where the stability test is rejected and the ML model is re-trained. 
	The horizontal red upper bound at 32 indicates the costly daily re-training of each borough strategy. Overall, the number of rejected monitoring tests are rarely close to this upper bound. 
	MLMA thus offers a
	flexible, automated way to smear out computing time over days since areas rarely require
	simultaneous re-training due to the heterogeneous market composition.
	In fact, the number of re-training instances per day is relatively concentrated and often below 10 until August 2020 after which it rises somewhat more regularly above 20. An explanation for this is that during the COVID pandemic, increased volatility in business activities made ML model forecast ability  more unstable, leading, in turn, to increased monitoring/re-training costs. This information generates actionable insights for better anticipating costs and work planning of the MLOps engineer. 
	
	Next, we summarize the borough-specific re-training instance results. Figure \ref{fig:break_information} (bottom panel) shows that the number of re-training instances per borough  is centered around 80 and, given its multi-modal shape, is highly variable among the boroughs. The duration, i.e.\ the number of days between subsequent re-training dates, is characterized by a long right tail. Indeed, while most of the boroughs require about weekly re-training, there are boroughs in London that are quite stable with re-training only every few weeks.  This clearly  calls for a borough-specific monitoring framework like MLMA rather than a periodic strategy that re-trains the ML model for all boroughs at exactly the same dates.

	\section{Conclusion} \label{Conclusion}
	Many businesses nowadays generate data streams in real time that are used to improve their own operations. This requires forecasts at granular levels, which can be made highly accurate thanks to the recent advances in machine learning (ML) models. The caveat is that successful ML models are computationally expensive to train. In particular, when they are running in production they require frequent re-training at a priori unknown dates
	because many business environments are inherently unstable causing forecast accuracy breakdowns. Hence, a machine learning operations (MLOps) monitoring procedure has to be put in place to decide when to re-train the models. 
	To this end, this paper introduces machine learning operations (MLOps) to the information systems literature.
	
	We focus on MLOps at digital platforms. In such business environments, continuous monitoring of when to re-train production-ready ML models is a labor intensive task  when  manually executed given their large amount of models in production.  The human monitoring cost is often considered too high and as a consequence platforms often revert to simple forecast methods; the precision of which is much lower than what is achievable with state-of-the-art ML models. The use of less accurate models has, however, direct negative business impact.  To solve this problem, we develop a new monitoring framework, the Machine Learning Monitoring Agent (MLMA), which drastically increases the productivity of MLOps engineers, thereby offering productivity gains while at the same time allowing to reach high forecast accuracies. Our framework keeps the MLOps engineer in the loop via key metrics so that a re-training decision can be based on historical data and real-time newly incoming information.  
	Methodology-wise, our monitoring framework signals when re-training of the algorithm is recommended because of a detected instability in a streaming forecast loss function. The process amounts to statistically testing if the forecast loss of a new incoming data batch significantly differs from a reference loss batch which dynamically adjusts over time. The implementation is straightforward and applicable to forecasts generated by any ML model.
	
	On-demand platforms naturally face high-frequency large-scale data stream settings subject to frequent changes, making them an appealing case study to demonstrate the value of our MLMA framework. We use a unique dataset consisting of 15-minute frequency demand for delivery streams in the 32 boroughs that constitute the London (UK) market.  In the evaluation period from  August 2019 to March 2021, everyday we produce random forest model based forecast streams for the sixty 15-minute bins of the next day. Their respective ground truth realizations allow computing loss streams; the mean of which is statistically compared with a reference loss batch. A significant difference alerts the MLOps engineer with several metrics and a re-training decision is made. The performance of MLMA is measured in three dimensions: Forecast accuracy, MLOps engineer monitoring cost,  and computing cost. We compare MLMA with several alternative monitoring strategies. It turns out that MLMA is able to obtain excellent forecast accuracy at a relatively low cost. This allows scaling MLMA to dimensions where accurate and manual monitoring are far too costly. While the excellent forecast performance under MLMA naturally comes with human and computational costs, our framework also allows gaining insights in time varying patterns in business data streams, to better understand how they interact with each other which is useful information that can be passed on to foster continuous model improvements within the MLOps architecture.
	
	We  deploy MLMA at an on-demand logistics platform, though it is important to note that our framework is general and has further potential in  a wide range of other application fields requiring monitoring high-frequency  forecast loss streams and detecting instability at large scale,  in a computationally feasible manner.
	E-commerce businesses and online market places such as Amazon or Walmart require fast forecasts  of
	bursty streaming web traffic data at high-frequency intervals that can rapidly adapt to structural changes, as part of their recommender systems to better understand consumer behavior and purchasing behavior. 
	Other data applications include, amongst others,
	forecasting intra-day residential electricity consumption data which have become available due to the increasing adoption of smart meters,
	forecasting tick-by-tick financial data which attracts growing attention from governmental regulators and industry,
	or
	forecasting streaming air-pollution data on a fine-resolution temporal and spatial scale to provide insights into air quality at local environments.
	MLMA offers exciting opportunities to facilitate the adoption and/or implementation of MLOps in these areas.
	
	\begingroup
	\setstretch{0.05}
	\linespread{0.5}
	\bibliographystyle{apalike}
	\bibliography{RomWil}
	\endgroup		
	\clearpage

	\newpage
	
	\renewcommand{\thetable}{A.\arabic{table}}
	\setcounter{table}{0}
	
	\renewcommand{\thefigure}{A.\arabic{figure}}
	\setcounter{figure}{0}
	
	\begin{appendices}
		
		\section{Monte Carlo Study} \label{app:MonteCarlo}
		The statistical monitoring test is repeatedly employed when new forecast loss batches become available. Given that the reference loss batch and therefore the null hypothesis changes when an accuracy breakdowns is detected,  we expect reasonable size properties of our test. 
		This is confirmed in Table \ref{tab:MonteCarlo} which presents Monte Carlo simulation results under the null hypothesis of forecast stability. For Gaussian data streams with lengths ranging between 10,000 and 100,000, and data batch sizes from 10 to 100 observations, the empirical rejection frequencies are only slightly higher than their nominal levels. For Chi-square (with five degrees of freedom) data streams the results are  similar.

		\begin{table}[ht]
			\begin{center}
				\caption{Rejection frequencies under forecast stability. \label{tab:MonteCarlo}}
				\begin{tabular}{lccccccc}
					\hline
					\multicolumn{8}{c}{Gaussian Data} \\[0.2cm]  
					Length stream  &  \multicolumn{3}{c}{5\% significance level} && \multicolumn{3}{c}{1\% significance level}  \\[0.1cm] \hline 
					&  \multicolumn{3}{c}{Batch size} && \multicolumn{3}{c}{Batch size}  \\[0.1cm] 
					& 10 & 50 & 100 &&  10 & 50 & 100  \\ 
					\cline{2-4}  \cline{6-8}\\[-0.4cm]
					
					10,000 &0.067	&	0.070	&	0.071	&&	0.013	&	0.013	&	0.013	\\
					20,000 &0.067	&	0.071	&	0.072	&&	0.013	&	0.013	&	0.013	\\
					50,000 & 0.067	&	0.071	&	0.072	&&	0.013	&	0.013	&	0.014	\\
					100,000 &0.067	&	0.071	&	0.072	&&	0.013	&	0.013	&	0.014	\\
					\hline
					\multicolumn{8}{c}{ } \\[-0.1cm]  
					\multicolumn{8}{c}{Chisquare(5) Data} \\[0.2cm]  
					Length stream  &  \multicolumn{3}{c}{5\% significance level} && \multicolumn{3}{c}{1\% significance level}  \\[0.1cm] \hline 
					&  \multicolumn{3}{c}{Batch size} && \multicolumn{3}{c}{Batch size}  \\[0.1cm] 
					& 10 & 50 & 100 &&  10 & 50 & 100  \\ 
					\cline{2-4}  \cline{6-8}\\[-0.4cm]
					
					10,000 &0.074	&	0.073	&	0.072	&&	0.022	&	0.016	&	0.015 \\
					20,000 &0.074	&	0.072	&	0.071	&&	0.022	&	0.016	&	0.015 \\
					50,000 &0.073	&	0.073	&	0.072	&&	0.022	&	0.016	&	0.015 \\
					100,000 &0.073	&	0.073	&	0.072	&&	0.022	&	0.017	&	0.015 \\
					\hline
				\end{tabular}
			\end{center}
			\raggedright	
			\footnotesize
			Notes: Rejection frequencies of the monitoring test under the null hypothesis of forecast stability. Batch size indicates the number of observations that are sequentially  added to the  stream.  The top (bottom) panel reports results for streams from Gaussian (Chi-square, with 5 degrees of freedom) data. 
			The results are based on 1,000 replications by running Algorithm \ref{alg:monitor}.  
		\end{table}
		
		In the case of homogeneous data streams modeled with linear regressions, \cite{White_ECA_1996} illustrate that standard tests reject too often.
		For our monitoring purposes, an oversized test implies more frequent re-training, at the benefit of forecast accuracy but the cost of computing time.  Our approach resembles \cite{Luo_JASA_2022_batches} who monitor abnormal data batches in data streams modeled with time invariant parametric functions. Using the  standard test of \cite{Hansen_Econometrica_1982}, their null hypothesis regards the equality of model parameters using score vectors between a fixed reference batch and a new incoming batch. In our setup, the forecast loss function streams are only temporarily in a stable regime, which explains why we define a new reference batch at each date with detected forecast instability.

		\renewcommand{\thetable}{B.\arabic{table}}
		\setcounter{table}{0}
		
		\renewcommand{\thefigure}{B.\arabic{figure}}
		\setcounter{figure}{0}
		
		\section{Platform Application: Additional Insights} \label{app:figures}
		We provide additional insights into the (i) typical intra-day and weekly patterns of the data streams, and (ii) the dynamics between the different boroughs.
		
\begin{figure}[H]
			\centering
			\includegraphics[width=0.49\textwidth]{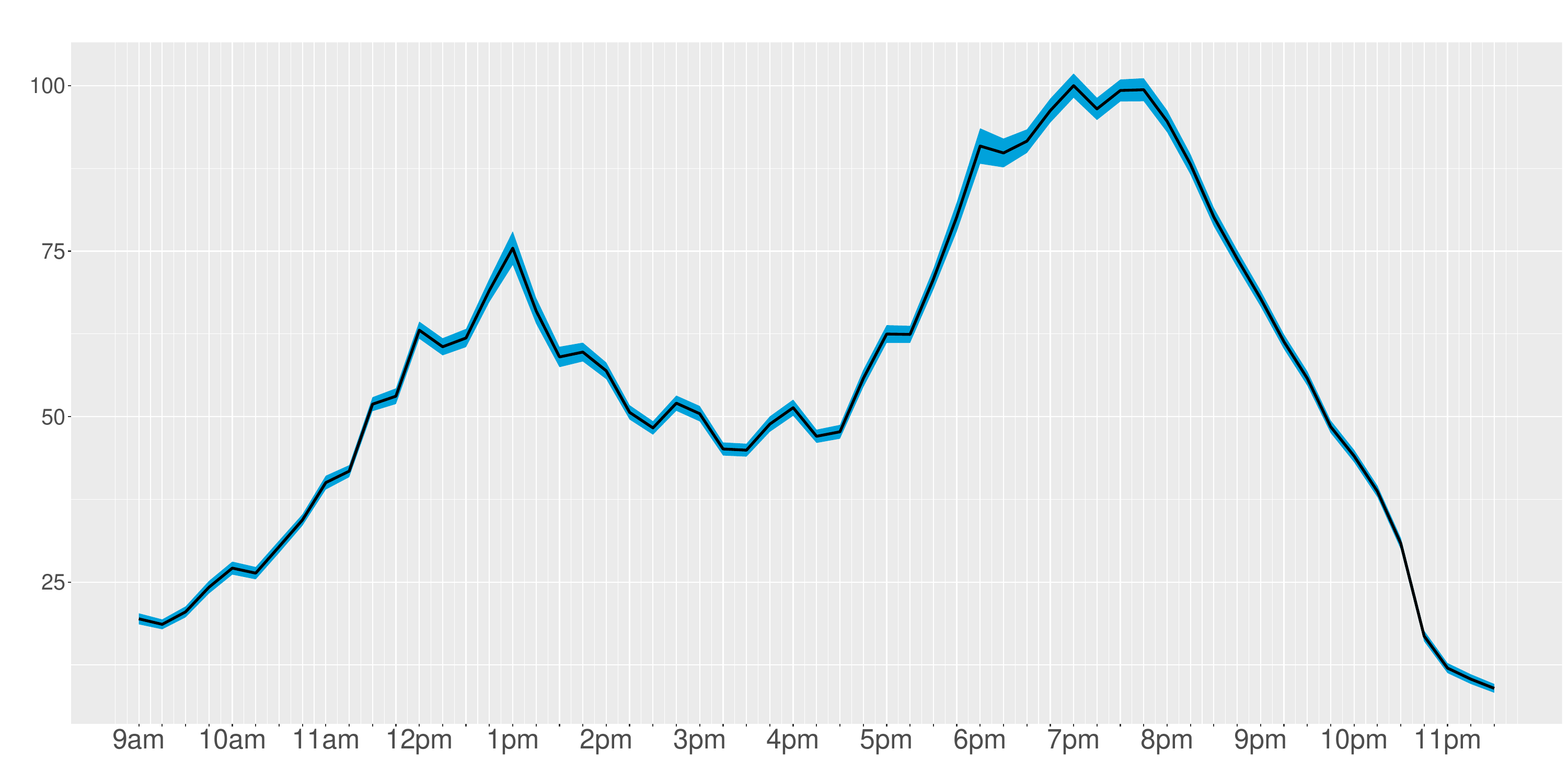}
			\includegraphics[width=0.49\textwidth]{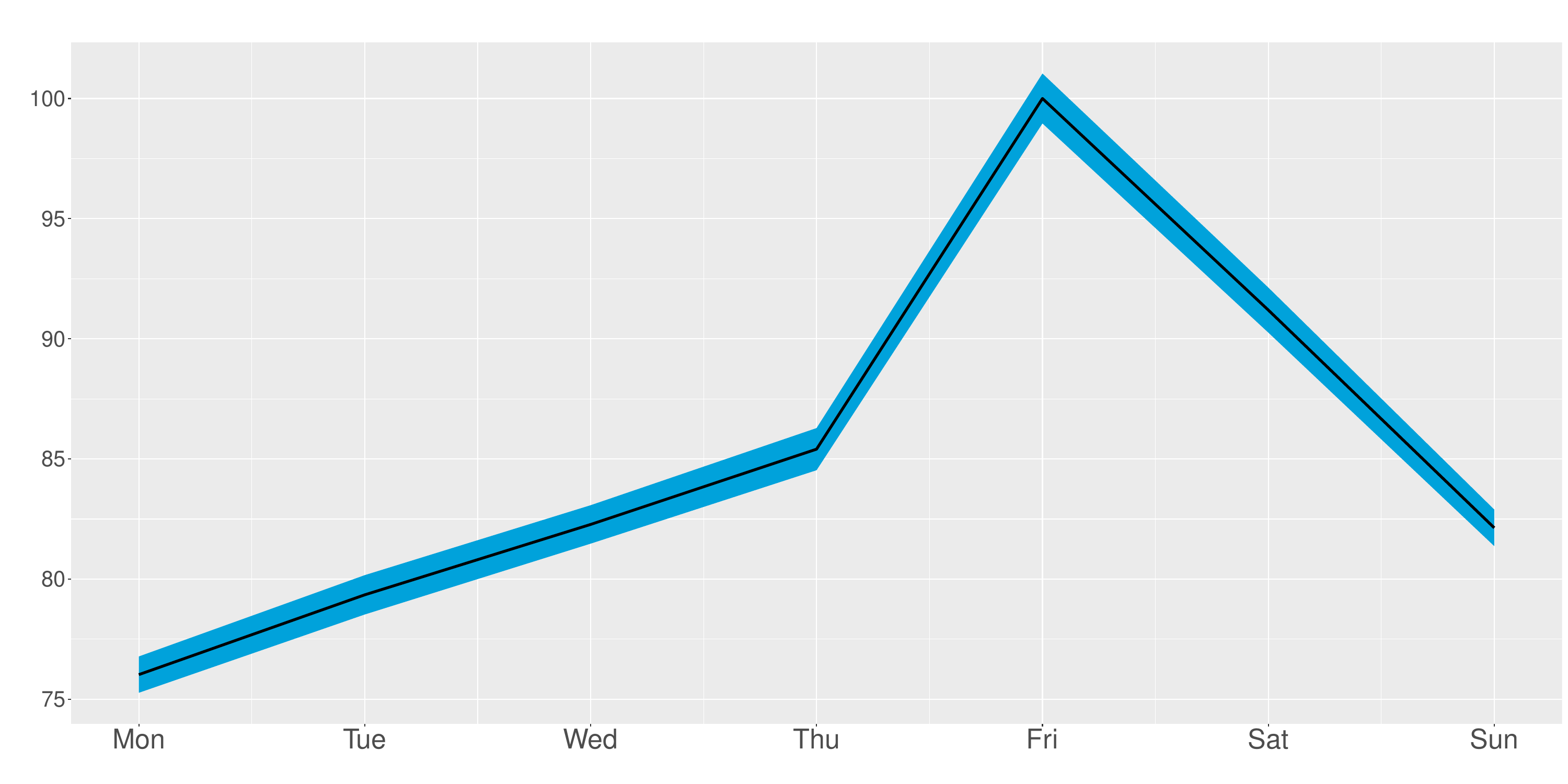}
			\caption{Intra-day 15-min average hourly demand  (left) and day of the week average demand  (right) for Tower Hamlets. One standard error bands are displayed in blue. \label{fig:TowerHamlets_seasonality}}
		\end{figure}

		Figure \ref{fig:TowerHamlets_seasonality} 
		(left)
		shows intra-day demand for Tower Hamlets. It becomes clear that the large majority of deliveries consists of food. In fact, demand starts low at 9am, first peaks around lunchtime between 12-2pm,  goes down only mildly between 2-5pm, but then sharply increases to peak around 7-8pm, after which it gradually decreases again to levels comparable to the morning hours. 
		Besides, pronounced day of the week fluctuations are present  as well, see Figure \ref{fig:TowerHamlets_seasonality} 
		(right).
		Average demand is lowest on Mondays, steadily increases until Thursdays, jumps on Fridays and Saturdays, and lowers on Sundays to typical Thursday levels.

		\begin{figure}[H]
			\centering
			\includegraphics[width=0.49\textwidth]{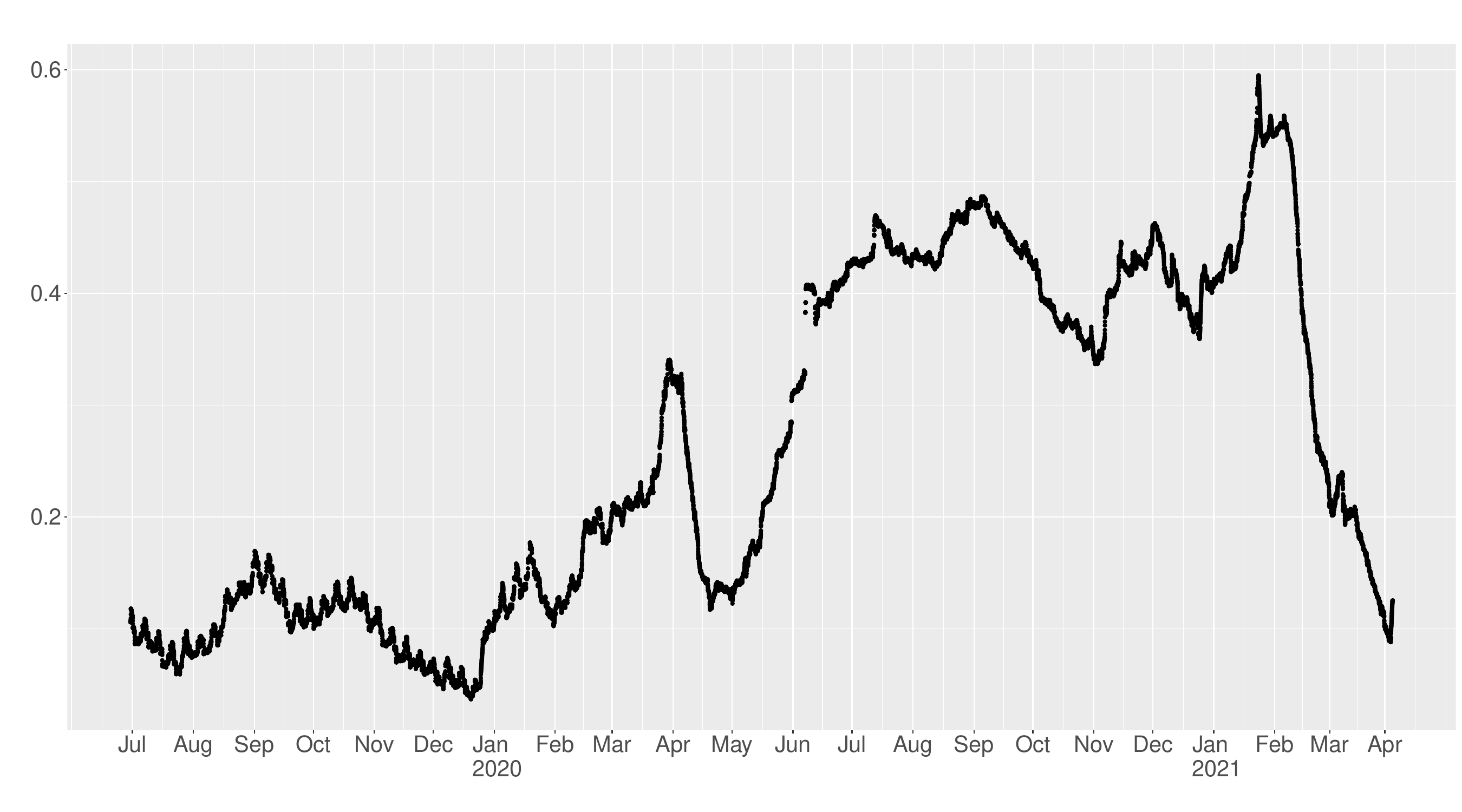}
			\includegraphics[width=0.49\textwidth]{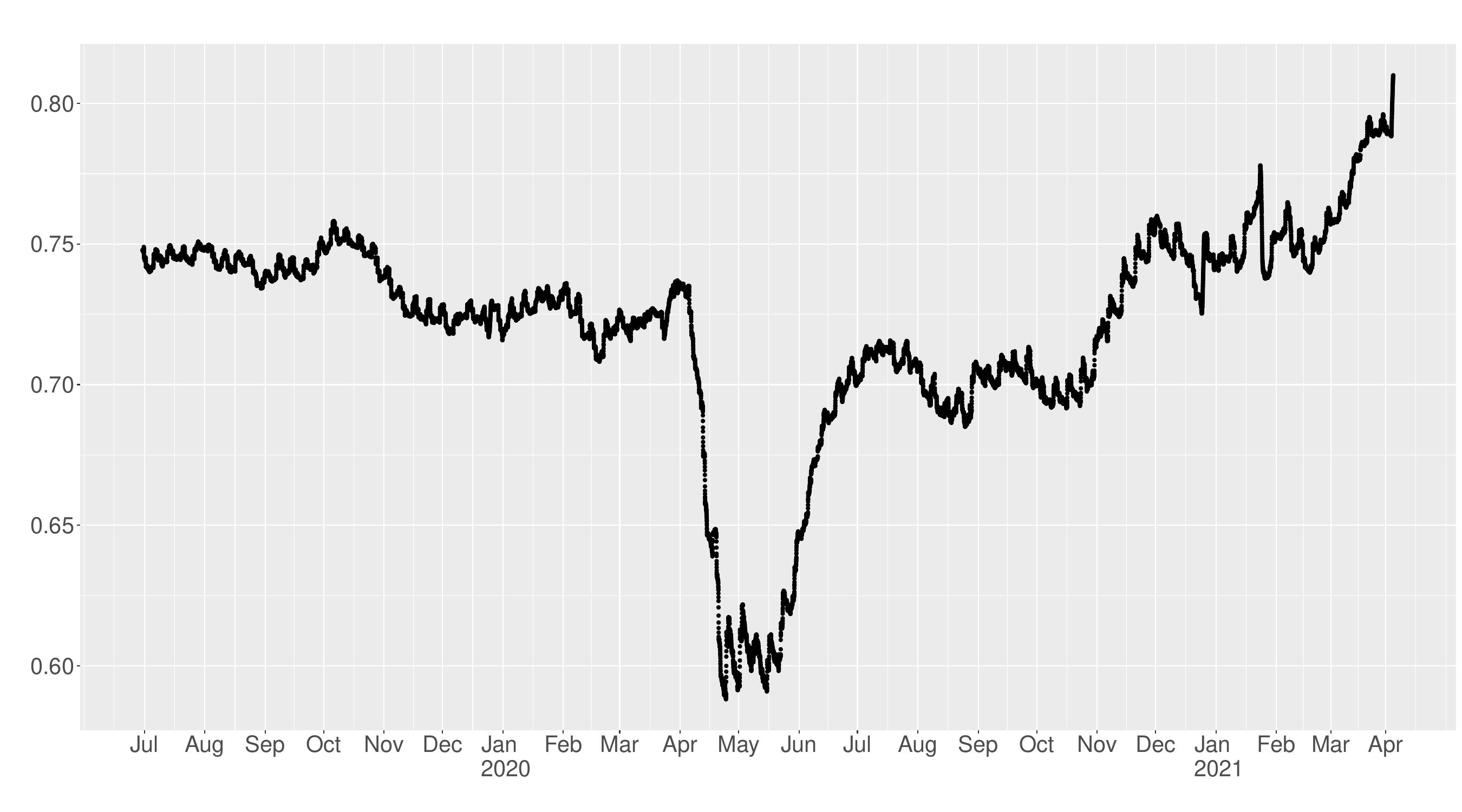}
			\caption{Streaming 15-min average correlations  for City of London (left)  and Barnet (right). \label{fig:tv_correlations}}
		\end{figure}

		The seasonality and trend  patterns are common to several boroughs which explains their relatively high average between borough correlation of 77\% (ranging between 16\% and 91\%) over the entire sample. Correlations, however, are strongly time-varying, as can be seen from Figure
		\ref{fig:tv_correlations} 
		which displays streaming 15-minute average correlations (for our forecast evaluation period) computed using demand from the thirty most recent days.  
		City of London  has specific idiosyncratic dynamics that makes it  least related to the other boroughs with average correlations dropping below 10\%, though it has also a relatively long spell late 2020 of average correlations hovering around 40\%. In contrast, Barnet, characterized by high demand, has high average streaming correlations around 75\% throughout the sample except mid 2020 when it fell to 60\% for more than a month. The time-varying nature of these correlations makes it hard to know in advance which boroughs are predictive for which other boroughs (if any) when building a forecast model.
		
		Finally, Table \ref{tab:SMAPE-per-borough} displays the SMAPE forecast performance of MLMA and the benchmarks (in the columns) for each of the 32 boroughs (in the rows). This table provides a disaggregate perspective on the overall results summarized in Table \ref{tab:SMAPE_all_boroughs}. 
		Tower Hamlets and Newham, reach average SMAPEs below 20. 
		In contrast, boroughs such as City of London and Sutton are more difficult to forecast and have average SMAPEs slightly above 30.

		\begin{table}[ht]
			\caption{SMAPE forecast performance per borough of MLMA and the benchmark monitoring methods. \label{tab:SMAPE-per-borough}}
			\resizebox{0.98\textwidth}{!}{\begin{minipage}{\textwidth}
					\centering
					\begin{tabular}{lcccccccccccc}
						\hline
						& MLMA &&&   \multicolumn{2}{c}{\underline{Do nothing}} && \multicolumn{2}{c}{\underline{Periodic}} && \multicolumn{1}{c}{\underline{On demand}}   \\
						&  &&& ML  & Naive  &&  Semester &   Daily &&   \\
						\hline
						TowerHamlets & 18.43 &&& 28.51 & 20.34 && 23.94 & 16.20 && 25.70  \\ 
						Wandsworth & 21.76 &&& 41.56 & 23.58 && 27.55 & 19.17 && 30.89  \\ 
						Camden & 22.55 &&& 43.50 & 25.77 && 28.76 & 20.66 && 29.68  \\ 
						Islington & 32.03 &&& 49.75 & 37.69 && 36.67 & 30.04 && 31.76  \\ 
						Westminster & 20.75 &&& 36.73 & 24.21 && 25.32 & 19.22 && 25.49  \\ 
						Lambeth & 22.35 &&& 31.26 & 25.08 && 24.54 & 20.02 && 24.74 \\ 
						City of London & 32.65 &&& 38.85 & 37.33 && 35.06 & 30.40 && 31.81 \\ 
						Kensington \& Chelsea & 25.67 &&& 37.53 & 29.51 && 29.67 & 24.24 && 26.98  \\ 
						Southwark & 22.34 &&& 41.02 & 24.78 && 27.51 & 20.07 && 33.25  \\ 
						Barking \& Dagenham & 31.49 &&& 47.82 & 37.73 && 39.18 & 29.03 && 32.38  \\ 
						Barnet & 21.65 &&& 43.20 & 22.70 && 27.63 & 18.60 && 26.38  \\ 
						Brent & 26.78 &&& 39.15 & 31.54 && 31.04 & 25.39 && 28.40  \\ 
						Ealing & 25.24 &&& 43.15 & 27.78 && 31.91 & 23.18 && 29.20  \\ 
						Greenwich & 22.40 &&& 42.38 & 23.88 && 28.06 & 19.94 && 25.84  \\ 
						Hackney & 28.19 &&& 42.63 & 31.88 && 29.88 & 26.36 && 28.40  \\ 
						Haringey & 21.97 &&& 43.06 & 24.20 && 29.73 & 19.74 && 28.71  \\ 
						Havering & 24.78 &&& 46.11 & 26.62 && 30.80 & 22.13 && 26.90  \\ 
						Hillingdon & 27.72 &&& 52.05 & 29.41 && 31.74 & 23.68 && 28.94  \\ 
						Kingston upon Thames & 26.52 &&& 42.78 & 29.08 && 30.15 & 24.17 && 28.50  \\ 
						Merton & 26.33 &&& 40.97 & 30.39 && 29.95 & 24.61 && 27.23  \\ 
						Newham & 19.06 &&& 33.27 & 21.31 && 24.50 & 17.28 && 25.58  \\ 
						Redbridge & 23.14 &&& 41.44 & 24.52 && 28.19 & 21.00 && 25.52  \\ 
						Lewisham & 22.15 &&& 40.84 & 24.19 && 26.08 & 20.05 && 27.46  \\ 
						Richmond upon Thames & 33.67 &&& 48.40 & 39.78 && 36.11 & 31.35 && 33.06  \\ 
						Hounslow & 27.20 &&& 47.47 & 28.78 && 34.07 & 22.83 && 27.20  \\ 
						Croydon & 22.25 &&& 44.01 & 23.29 && 29.90 & 19.59 && 30.57  \\ 
						Enfield & 23.18 &&& 42.29 & 24.90 && 30.46 & 20.40 && 27.45  \\ 
						WalthamForest & 22.81 &&& 41.06 & 25.31 && 28.85 & 20.59 && 26.82 \\ 
						Harrow & 26.29 &&& 43.74 & 29.93 && 30.65 & 23.96 && 27.92  \\ 
						Bromley & 29.24 &&& 48.73 & 31.20 && 33.56 & 25.37 && 30.76  \\ 
						Sutton & 34.24 &&& 49.53 & 39.06 && 38.71 & 31.06 && 32.92  \\ 
						Bexley & 26.68 &&& 43.38 & 29.08 && 35.34 & 24.31 && 29.73  \\ 
						\hline
					\end{tabular}
			\end{minipage}}
		\end{table}

	\end{appendices}
\end{document}